\documentclass[twocolumn,astrosymb,tighten,twocolappendix]{aastex701}

\usepackage{amsmath}
\usepackage{amssymb}
\usepackage{hyperref}
\usepackage{nameref}

\begin{document}

\title{Probing the Reionization History and Bubble Sizes with JWST Lyman-$\alpha$ Fraction Measurements}

\author[0000-0001-8746-4753]{Aritra Kundu}
\affiliation{Department of Physics \& Astronomy, University of Pennsylvania, 209 S 33rd St., Philadelphia, PA 19104, USA}
\email[show]{aritra@sas.upenn.edu}  

\author[0000-0002-3950-9598]{Adam Lidz}
\affiliation{Department of Physics \& Astronomy, University of Pennsylvania, 209 S 33rd St., Philadelphia, PA 19104, USA}
\email[]{alidz@sas.upenn.edu}

\author[0000-0003-4070-497X]{Guochao Sun}
\affiliation{CIERA and Department of Physics and Astronomy, Northwestern University, 1800 Sherman Ave, Evanston, IL 60201, USA}
\email[]{guochao.sun@northwestern.edu}

\begin{abstract}

Recent James Webb Space Telescope (JWST) measurements quantify the fraction of ultraviolet-selected galaxies showing detectable Lyman-alpha (Ly-$\alpha$) lines at $z \sim 6-14$. These Ly-$\alpha$ fractions are partly shaped by Ly-$\alpha$ scattering from neutral hydrogen in the surrounding intergalactic medium (IGM), and so the observed Ly-$\alpha$ fraction evolution can inform our understanding of the reionization history and the size distribution of ionized bubbles during reionization. We compare the JWST measurements with Ly-$\alpha$ transmission calculations based on both the \textsc{Thesan} reionization simulations and a flexible semi-analytic model. In each case, we calibrate a sub-grid model for the intrinsic Ly-$\alpha$ emission lines emerging from the interstellar medium (ISM) and circumgalactic medium (CGM) of the host galaxies using empirical estimates of the Ly-$\alpha$ equivalent width (EW) distributions at $z \sim 6$. We find that the reionization history and bubble sizes in  \textsc{Thesan} yield good agreement with current JWST Ly-$\alpha$ fraction observations, with a reduced $\chi^2_\nu = 0.4-1.2$. Our semi-analytic models and \textsc{Thesan} Ly-$\alpha$ transmission calculations agree well, allowing us to explore a wide range of reionization models. 
We derive $1\sigma$ bounds on the average mass-weighted neutral fractions of: $\langle x_{\mathrm{HI}} \rangle$ = $0.38^{+0.14}_{-0.38}$ ($0.52^{+0.09}_{-0.21}$), $0.57^{+0.23}_{-0.57}$ ($0.68^{+0.12}_{-0.39}$), and $0.93^{+0.05}_{-0.17}$ ($0.89^{+0.11}_{-0.15}$) at $z \sim$ 7, 8, and 11, based on the Ly-$\alpha$ fraction measurements with EW $>25\,\mathrm{\AA}$ ($>10\,\mathrm{\AA}$).

\end{abstract}



\section{Introduction}
\label{sec:intro}

Recent James Webb Space Telescope (JWST) measurements extend estimates of the fraction of ultraviolet (UV)-selected galaxies which emit prominent Lyman-alpha (Ly-$\alpha$) lines out to $z \sim 14$ \citep{Curtis-Lake_2023, Hsiao_2024, Umeda_2024, Nakane_2024, Tang_2024a, Tang_2024b, Jones_2024, Jones_2025, Kageura_2025, Witstok_2025, Mason_2026, Napolitano_2026b}. These Ly-$\alpha$ fraction measurements trace, in part, the reionization history of the universe: as the intergalactic medium (IGM) grows more neutral towards high redshift, Ly-$\alpha$ photons are increasingly likely to scatter out of the line-of-sight, suppressing the Ly-$\alpha$ fraction. The unprecedented lever arm in redshift from the new JWST observations enables constraints on most of the reionization history, including early phases of this process at $z \sim 14$ through to its completion around $z \sim 5.5-6$. The Ly-$\alpha$ fractions are also sensitive to the patchy spatial structure of reionization, encoding valuable information regarding the size distribution of the ionized bubbles around UV-selected galaxies.
Here, we present theoretical models to help interpret current JWST Ly-$\alpha$ fraction estimates and to prepare for more precise future measurements. 

To place the Ly-$\alpha$ fraction measurements in context, the Epoch of Reionization (EoR) is a phase transition during which hydrogen in the IGM transforms from largely neutral to mostly ionized, driven by UV radiation from the first stars, galaxies, and accreting black holes \citep{Barkana_2001, Furlanetto_2006a}. 
Current observational evidence suggests that star-forming galaxies are the main drivers of reionization \citep{Shapiro_1994, Madau_1999, Faucher-Giguere_2008, Becker_2013, Jiang_2025}, with the process completing between $z \sim 5.5-6$ \citep{Becker_2001, Fan_2006, Mortlock_2011, Schroeder_2013, McGreer_2015, Becker_2015, Greig_2017, Eilers_2019, Wang_2020, Zhu_2021, Bosman_2022, Zhu_2023,Zhu_2024,Spina_2024}.  
Initial estimates showed a rapid decline in the Ly-$\alpha$ fractions between $z=6$ and $z=7$ \citep{Schenker_2014,Tilvi_2014, Pentericci_2018, Mason_2019, Jung_2020}, while JWST is improving the sample sizes and redshift coverage of these measurements \citep{Kageura_2025, Napolitano_2026a, Napolitano_2026b}. 

Ly-$\alpha$ emission lines have long been recognized as a powerful tracer of high-redshift galaxy populations \citep{Partridge_1967}. 
Ly-$\alpha$ photons are produced primarily during recombination cascades when
an electron transitions from the first excited state to the ground state in a hydrogen atom. These photons subsequently undergo resonant scattering 
off neutral hydrogen in the interstellar medium (ISM) and circumgalactic medium (CGM) of the host galaxy, and are further attenuated by neutral hydrogen in the surrounding IGM \citep{Santos_2004, Dijkstra_2007, Laursen_2011, Dijkstra_2014, Gronke_2021, Park_2021, Smith_2022}. 
The Ly-$\alpha$ line profile emerging from the ISM/CGM is shaped by the distribution and kinematics of hydrogen in the host galaxy and the surrounding dark matter halo. 
Photons that are initially blue-ward of line center will redshift into the Ly-$\alpha$ resonance in the IGM, where there is a high probability for scattering out of the line of sight, even if the surrounding IGM has only a small residual neutral fraction \citep{Gunn_Peterson_1965}. On the other hand, red-side photons shift away from resonance as they propagate, and are noticeably attenuated by the damping wing (DW) of the Ly-$\alpha$ line, due to its natural linewidth, only when the IGM retains a significant neutral fraction \citep{Miralda-Escude_1998}. 
Therefore, one generally expects Ly-$\alpha$ lines to become increasingly attenuated as sources are observed at progressively earlier stages in the EoR due mainly to the growing DW optical depth. However, the precise evolution is challenging to predict as it depends on the distribution and kinematics of neutral hydrogen in the ISM and CGM, and on the morphology of remaining neutral gas in the IGM.

Motivated by these challenges, we combine an empirically-calibrated Ly-$\alpha$ line profile and Ly-$\alpha$ escape fraction model to account for ISM/CGM scattering, with a simulation from the \textsc{Thesan} project \citep{Kannan_2022, Garaldi_2022, Garaldi_2024, Smith_2022} to capture scattering in the IGM. The \textsc{Thesan} simulations are state-of-the-art radiation-hydrodynamic 
simulations of reionization, providing detailed models for the reionization history and the size distribution of ionized bubbles at different stages of the reionization process. 

An empirical calibration is necessary, however, to account for poorly-resolved ISM and CGM scales. We employ a flexible parameterization for the shape of the Ly-$\alpha$ line profile emerging after ISM/CGM scattering and dust attenuation, allowing for outflows which may imprint a red-ward shift upon escaping Ly-$\alpha$ photons and influence subsequent IGM scattering. 
The distribution of escaping Ly-$\alpha$ luminosities is anchored to empirical estimates of the probability distribution of Ly-$\alpha$ equivalent widths (EWs), conditioned on a galaxy's UV luminosity, near the end of reionization \citep{Tang_2024a,Tang_2024b}. We then use \textsc{Thesan} simulation outputs to calculate the scattering in the IGM at higher reionization-era redshifts. Throughout, we refer to the Ly-$\alpha$ line profiles and luminosities leaving the ISM/CGM, before encountering the IGM, as ``intrinsic''. 
Our fundamental simplifying assumption is that the {\em intrinsic} conditional EW distributions do not evolve with redshift, and so the observed decrease in Ly-$\alpha$ fractions between $z \sim 6-14$ is primarily driven by scattering in an increasingly neutral IGM (e.g. \citealt{Mason_2018}).

We also demonstrate that a simplified semi-analytic model reproduces the main features of the IGM DW calculations in \textsc{Thesan}. The semi-analytic model assumes that each UV-emitting galaxy is surrounded by an ionized bubble, of specified size, with a spatially-uniform neutral fraction outside of the bubble. This description allows analytic calculations of the DW optical depths, but neglects spatial variations in the neutral fraction outside of the local ionized bubble. Nevertheless, we find that this provides a good approximation for calculating the distribution of Ly-$\alpha$ transmission coefficients, which are shaped mainly by the optical depths relatively close to line center.

Our work overlaps with the recent study of \cite{Neyer_2025}, who also use the \textsc{Thesan} simulations to help interpret recent Ly-$\alpha$ fraction measurements. As we will discuss, our work differs from this study primarily in our different approaches for calibrating the intrinsic Ly-$\alpha$ line properties. Our flexible semi-analytic description also allows us to explore variations around the \textsc{Thesan} reionization model. Similar analyses were also carried out by \citet{Mason_2018, Kageura_2025, Mason_2026} to constrain the average neutral fraction of hydrogen during the EoR, using \textsf{21cmFAST} semi-numerical simulations \citep{Mesinger_2011}. We compare with those earlier studies in what follows. 

This paper is organized as follows. In Section \ref{subsec:thesan}, we overview relevant aspects of the \textsc{Thesan} simulations. Section \ref{subsec:EW_calibration} describes our EW calibration procedure and Section \ref{subsec:Lya_model} discusses our Ly-$\alpha$ emission line profile and DW optical depth modeling. 
Section \ref{subsec:bubble_sizes} characterizes the bubble size distributions, conditioned on UV luminosity, at different stages of reionization in the \textsc{Thesan} simulations. Section \ref{sec:results} presents our main results, comparing the resulting predictions for the Ly-$\alpha$ fraction evolution with current measurements.Finally, we conclude in Section \ref{sec:conclusions}.

\section{Methods} 
\label{subsec:methods}

\subsection{\textsc{Thesan} Simulations} 
\label{subsec:thesan}

The \textsc{Thesan} project is a suite of state-of-the-art radiation-magneto-hydrodynamic (RMHD) cosmological simulations \citep{Kannan_2022, Garaldi_2022, Garaldi_2024, Smith_2022} that self-consistently model reionization and the galaxies responsible for it, while using sub-grid prescriptions based on those calibrated in IllustrisTNG \citep{Weinberger_2017, Pillepich_2018}.  These simulations allow us to model Ly-$\alpha$ scattering, and to characterize the bubble size distributions around simulated UV-selected galaxies. 
Each \textsc{Thesan} simulation has a comoving box length of $L_{\rm box}$ = 95.5 cMpc, and adopts the following cosmological parameters based on \citet{Planck_2016} measurements: $h = 0.6774$, $\Omega_m = 0.3089$, $\Omega_\Lambda = 0.6911$, $\Omega_b = 0.0486$, $n_{\rm s} = 0.9667$, and $\sigma_8 = 0.8159$. The same parameter values are adopted throughout our calculations. We use the high-resolution, fiducial simulation, \textsc{Thesan-1}, which tracks 2100$^3$ dark matter and gas particles each, corresponding to particle masses of $m_{\rm DM}$ = 3.12 $\times$ 10$^{6}$ $M_\odot$ and $m_{\rm gas}$ = 5.82 $\times$ 10$^{5}$ $M_\odot$, respectively. The minimum total mass of resolved dark matter halos is $M_{\rm halo} \sim 10^8 \, h^{-1}$ $M_\odot$,
while the simulation is run down to $z \sim 6$. In this work, we use the \textsc{Thesan} galaxy catalogs and gridded estimates of the mass-weighted neutral hydrogen fraction fields from each simulation snapshot, tabulated
on Cartesian grids with 512$^3$ cells.

\subsection{Equivalent Width Distribution Calibration at $z \sim 6$} \label{subsec:EW_calibration}

The Ly-$\alpha$ EW characterizes the observed Ly-$\alpha$ emission line strength relative to the UV continuum \citep{Dijkstra_2012}:

\begin{equation}
    \mathrm{EW}_{\mathrm{obs}} = \frac{L_{\alpha,\mathrm{obs}}}{L_{\rm UV, \lambda}(\lambda_\alpha)} = \frac{\lambda_\alpha}{\nu_\alpha} \left(\frac{1500\, \rm{\AA}}{1216\, \rm{\AA}}\right)^{\beta + 2} \frac{L_{\alpha,\mathrm{obs}}}{L_{\mathrm{UV},\nu}},
    \label{eq:ew_def}
\end{equation}

where $L_{\mathrm{UV},\nu}$ is the specific continuum luminosity in units of $\mathrm{erg \, s^{-1} \, Hz^{-1}}$, $L_{\rm UV, \lambda}(\lambda_\alpha)$ is the specific continuum luminosity at the Ly-$\alpha$ rest-frame wavelength in units of $\mathrm{erg \, s^{-1} \, \rm{\AA}^{-1}}$, $L_{\alpha,\rm{obs}}$ is the total observed Ly-$\alpha$ luminosity in units of $\mathrm{erg \, s^{-1}}$, while
$\lambda_\alpha = 1216\, \rm{\AA}$ and $\nu_\alpha = 2.47 \times 10^{15}$ Hz are the rest-frame wavelength and frequency of the Ly-$\alpha$ line, respectively. The continuum normalization $L_{\mathrm{UV}, \nu}$ is given at
a rest-frame wavelength of $1500\, \rm{\AA}$, while the continuum spectrum is assumed to scale as $L_{\mathrm{UV},\nu} \propto \nu^{-\beta -2}$, with $\beta \sim -2$. The EW is specified here in units of $\rm{\AA}$, and in the rest frame of the host galaxy.\footnote{Note that the notation $\mathrm{EW}_{\mathrm{obs}}$ means the EW after IGM attenuation, even though this is a rest-frame EW as opposed to an EW in the observer's frame. We use rest frame EWs throughout.}

Ultimately, our goal is to model the fraction of UV-selected galaxies whose observed EWs in different UV magnitude bins exceed certain thresholds, as a function of redshift. As alluded to in Section~\ref{sec:intro} 
we consider the observed Ly-$\alpha$ luminosity and EW to be a product of an intrinsic EW (and/or luminosity) with an IGM transmission factor:
\begin{equation}
\mathrm{EW}_{\mathrm{obs}} = \mathrm{EW_{in}} \times \mathcal{T_\alpha}, 
\label{eq:ew_trans}
\end{equation}
where $\mathrm{EW_{in}}$ accounts for Ly-$\alpha$ scattering and dust attenuation within the poorly-resolved ISM and CGM, while the transmission through the IGM, $\mathcal{T_\alpha}$, can be robustly modeled with \textsc{Thesan}.

We then calibrate the distribution of intrinsic EWs based on empirical measurements near the end of reionization at $z \sim 6$. Subsequently, we predict the distribution of observed EWs at higher redshifts by modeling the transmission evolution into the EoR. As detailed in Appendix \ref{append_sec:EW_dist}, the observed EW distribution among galaxies with a given UV magnitude, $M_{\rm UV}$, can be described by a convolution:

\begin{align}
    \mathrm{P}_{z}(\mathrm{EW}_{\mathrm{obs}} | M_{\mathrm{UV, z}}) = &
 \int_0^1 d\mathcal{T}_{\alpha, z} \, \frac{1}{\mathcal{T}_{\alpha, z}} \, \mathrm{P}_{\alpha}(\mathcal{T}_{\alpha, z}|M_{\mathrm{UV}, z}) \nonumber \\
    & \times \mathrm{P}_{\rm in}\left(\mathrm{EW_{in}} = \frac{\mathrm{EW}_{\mathrm{obs}}}{\mathcal{T}_{\alpha, z}} \,\middle\vert\, M_{\mathrm{UV}} \right).
    \label{eq:EW_dist_cond}
\end{align}

Here, $\mathrm{P}_\alpha(\mathcal{T}_{\alpha, z}|M_{\mathrm{UV}, z})$ is the conditional IGM transmission distribution at redshift $z$ among galaxies with UV magnitude $M_{\mathrm{UV}}$, while $\mathrm{P}_{\rm in}(\mathrm{EW_{in}}|M_{\mathrm{UV}})$ is the conditional distribution of {\em intrinsic} EWs. Our fundamental assumption is that $\mathrm{P}_{\rm in}(\mathrm{EW_{in}}|M_{\mathrm{UV}})$ does not itself evolve across the reionization-era redshifts of interest. Under this commonly-adopted model \citep{Dijkstra_2012, Dijkstra_2014, Mason_2018, Tang_2024a, Tang_2024b}, the redshift evolution of the observed EW distribution is driven solely by 
changes in the conditional transmission distribution during reionization. 

In practice, $\mathrm{P}_{\rm in}(\mathrm{EW}_{\rm in}|M_{\rm UV})$ is not directly observable, but may be calibrated empirically. Specifically, to infer it from observations near the end of reionization we need to account for both IGM scattering on the blue-side of the Ly-$\alpha$ line, where the opacity is large even after reionization completes, and for any red-side IGM DW absorption close to reionization's completion. Here, we anchor to the observed conditional EW distributions at $z \sim 6$ from \citet{Tang_2024b}, while accounting for the conditional transmission distributions at this redshift from \textsc{Thesan}. In Appendix \ref{append_sec:EW_dist} we show that the intrinsic distribution may be inferred as:
\begin{align}
    \mathrm{P_{in}}(\mathrm{EW}_{\rm{in}}|M_{\mathrm{UV}}) 
    &= \int_0^1 d\mathcal{T}_{\alpha, 6} \, \mathcal{T}_{\alpha, 6} \, \mathrm{P}_{\alpha}(\mathcal{T}_{\alpha, 6} | M_{\mathrm{UV, 6}}) \nonumber \\
    &\hspace{1.25cm} \times \mathrm{P}_{6}(\mathrm{EW_{\rm{in}}\mathcal{T}_{\alpha, 6}} | M_{\mathrm{UV, 6}}),
    \label{eq:EW_in_infer}
\end{align}
where $\mathrm{P}_{\alpha}(\mathcal{T}_{\alpha, 6} | M_{\mathrm{UV, 6}})$ is the $z=6$ IGM conditional transmission distribution determined from \textsc{Thesan}, and $\mathrm{P}_{6}(\mathrm{EW_{\rm{in}}\mathcal{T}_{\alpha, 6}} | M_{\mathrm{UV, 6}})$ is the observed EW distribution at $z=6$. In practice, there is relatively little IGM DW absorption near the end of reionization and so $\mathrm{P}_{\alpha}(\mathcal{T}_{\alpha, 6} | M_{\mathrm{UV, 6}})$ is generally a sharply-peaked distribution. 

The observed EW distribution at $z=6$ is assumed to follow the lognormal fits from \citet{Tang_2024b}:
\begin{align}
\mathrm{P}_6(\mathrm{EW}_{\mathrm{obs}}|M_{\rm UV, 6}) = & \frac{1}{\sqrt{2 \pi \sigma^2}}\frac{1}{\mathrm{EW}_{\mathrm{obs}}} \nonumber \\ 
& \times \exp\left[-\frac{(\ln \mathrm{EW}_{\mathrm{obs}} - \mu)^2}{2\sigma^2}\right], 
\label{eq:psix_logn}
\end{align}
where the best-fit parameters and uncertainties are given in Table~\ref{tab:tang24} for reference. See \citet{Tang_2024b} for a comparison between their lognormal fits and alternative parameterizations \citep{Treu_2012,DeBarros_2017,Mason_2018}. For comparison with Ly-$\alpha$ fraction measurements in the current literature, we will focus on a UV-magnitude bin with $-20.25 < M_{\rm UV} < -18.75$, and hence we use the results from \citet{Tang_2024a} in the same bin.

We account for dust attenuation of the intrinsic UV luminosities among our simulated galaxies using the empirical model in \citet{Neyer_2025}. This is based on fitting a redshift-dependent Infrared Excess-Ultraviolet relationship to ALMA observations at $z \sim 4-7$ \citep{Bouwens_2016}, with fitting parameters reported in \citet{Kannan_2023}. The attenuation decreases with redshift and is larger for intrinsically UV-bright galaxies. For example, in the case of an intrinsic UV luminosity of $M_{\rm UV} =20 \, (-18)$, the observed luminosity is $M_{\rm UV} = -19.5\, (-17.8)$ at $z = 6$, while it is $M_{\rm UV} = -19.7\, (-17.9)$ at $z=14$. Hence, the dust corrections are fairly small for the regimes of interest in this work.

In summary, Eqns.~\ref{eq:ew_trans}-\ref{eq:psix_logn} specify our approach for calibrating to the observed conditional EW distributions near the end of reionization. In order to determine how the EW distributions evolve into the EoR, we need next to model the conditional transmission distributions and their redshift dependence. 

\begin{deluxetable*}{ccc}
\tablewidth{0pt} 
\tablecaption{Equivalent Width Distribution Parameters \label{tab:tang24}}
\tablehead{
\colhead{Sample} & \colhead{$e^{\mu}$ ($\rm{\AA}$)} & \colhead{$\sigma$}}

\startdata 
$z \sim 6$, $-20.25 < M_{\rm{UV}} < -18.75$ & $8^{+4}_{-3}$ & $1.85^{+0.42}_{-0.33}$ \\
\enddata
\tablecomments{Best-fit parameters, and their $1\sigma$ uncertainties for the conditional EW distribution model
of Eqn.~\ref{eq:psix_logn}, based on the values from Table 2 in \citet{Tang_2024b}. We note a typo in that work. The widths ($\sigma$) of the distributions are labeled as being in dex, but they in fact refer to the $\sigma$ of $\ln (\mathrm{EW}_{\mathrm{obs}})$ and so are not actually in dex.}
\end{deluxetable*}

\subsection{Modeling Ly-$\alpha$ Emission Lines and IGM Scattering}
\label{subsec:Lya_model}

Our models for the Ly-$\alpha$ emission lines from UV-selected galaxies require two additional ingredients. First, we need to model the intrinsic line profiles emerging from the ISM and CGM. Second, we discuss our calculations of Ly-$\alpha$ scattering from intergalactic neutral hydrogen.

\subsubsection{Intrinsic Ly-$\alpha$ Line Profiles}
\label{subsubsec:Ly_a_luminosity}

We assume a Gaussian intrinsic line profile
\citep{Dijkstra_2007, Zheng_2010, Laursen_2011, Mason_2018}:

\begin{equation}
    L_{\alpha, v}^{\rm int} = \frac{L_{\alpha}^{\rm int}}{\sqrt{2 \pi \sigma_v^2}} \, \exp\!\left[-\frac{(\Delta v - 2v_{\rm out})^2}{2\sigma_v^2}\right],
    \label{eqn:Lya_profile}
\end{equation}
where $\Delta v$ is the velocity relative to the galaxy's systemic redshift, $\sigma_v$ is the linewidth, and $v_{\rm out}$ is an outflow speed. Here, $L_{\alpha, v}^{\rm int}$ is the luminosity per unit velocity interval, while the total integrated line luminosity is $L_{\alpha}^{\rm int}$. Note that our results are independent of the line luminosity normalization in this equation since we anchor to the observed EW distributions. The shape of the line profile is, however, important for determining the IGM transmission coefficient, $\mathcal{T}_\alpha$. 

We model outflow effects using a shell-model description in which interstellar neutral hydrogen is concentrated in an expanding spherical shell \citep{Verhamme_2006, Dijkstra_2017}. The Ly-$\alpha$ photons produced within central HII regions scatter upon reaching the shell. Specifically, the photons received by an observer are mainly back-scattered by the expanding shell, with redshifts imprinted upon entering the shell's frame and when photons re-enter the observer's frame after back-scattering, yielding a net shift of $\Delta \nu/\nu = - 2 v_{\rm out}/c$. 
The outflow velocity is expected to scale with the depth of the host galaxy's gravitational potential well \citep{Murray_2005,Oppenheimer_2006,Chisholm_2015}. We assume that $v_{\rm out} = a v_{\rm circ}$, where $v_{\rm circ}$ is the circular velocity of the host dark matter halo at the virial radius and $a$ is a proportionality constant. 
The circular velocity is related to the host dark matter halo mass $M_{\rm halo}$ according to \citep{Barkana_2001}:
\begin{equation}
    v_{\rm circ} = 82 \, \left(\frac{M_{\rm halo}}{10^{10} \,M_\odot}\right)^{1/3} \left(\frac{1+z}{10}\right)^{1/2} \, \rm km \, s^{-1}. 
    \label{eqn:v_circ}
\end{equation}
We consider $a=0$ and $1$ to gauge the importance of outflows. We further adopt $\sigma_v = v_{\rm circ}$ throughout, since the linewidths may be determined by dynamical Doppler broadening with a characteristic velocity set by the circular velocity at the halo virial radius \citep{Santos_2004, Dijkstra_2007, McQuinn_2007b}. For example, the linewidths may reflect turbulent motions in the dispersion-dominated gas of these high-redshift galaxies \citep{Sun_26}. 
Note that Ly-$\alpha$ scattering can also broaden the emerging lines, with the broadening depending on the column density of neutral hydrogen in the ISM/CGM of the host galaxy. We find similar results, however, upon adopting the alternate emission line profile in \citet{Neyer_2025}, which incorporates broadening from scattering, as discussed in Section \ref{sec:comp_other_work}.

\subsubsection{Damping Wing Optical Depths and Transmission Coefficients}
\label{subsec:DW_transmission}

We calculate the IGM transmission using the line profile of Eqn.~\ref{eqn:Lya_profile}. We assume that flux blue-ward of Ly-$\alpha$ at the systemic redshift is completely absorbed, while we approximate the absorption profile red-ward of Ly-$\alpha$ as being due entirely to the DW of the line. As mentioned in the Section \ref{sec:intro}, the blue-side photons shift into resonance where even small residual neutral fractions lead to large optical depths and nearly complete absorption, as is evident from Ly-$\alpha$ forest observations near $z \sim 6$ \citep{Fan_2006, Yang_2020}.
In principle, infalling gas can also lead to resonant absorption red-ward of line center \citep{Dijkstra_2007}, but we neglect this possibility (and ignore any other peculiar velocity effects) in what follows for simplicity.

Under these approximations, the Ly-$\alpha$ transmission coefficient may be calculated as:
\begin{equation}
    \mathcal{T_\alpha} = \frac{\int_0^\infty d\nu L_{\alpha, \nu}^{\rm int} \, \rm e^{-\tau_{\rm DW}(\nu)}}{\int_{-\infty}^{\infty} d\nu L_{\alpha, \nu}^{\rm int}},
    \label{eqn:transmission}
\end{equation}
where the integral is over the intrinsic line profile of Eqn.~\ref{eqn:Lya_profile}, except expressed here in terms of frequency rather than velocity offsets using $|\Delta \nu/\nu| = \Delta v/c$. The numerator covers only frequencies red-ward of the systemic redshift, while the denominator includes the full line. That is, frequencies blue-ward of the systemic redshift are assumed to be entirely absorbed while the red-side transmission is governed by the DW optical depth, $\tau_{\rm DW}(\nu)$. Note that under these approximations, in the absence of outflows, $\mathcal{T_\alpha} \rightarrow 1/2$ as $\tau_{\rm DW}(\nu) \rightarrow 0$. This occurs because exactly $1/2$ of the Gaussian line profile (Eqn.~\ref{eqn:Lya_profile}) shifts into resonance and scatters out of the line-of-sight in the absence of outflows and DW absorption. Outflows increase the transmission coefficients, as we will discuss. 

We calculate the Ly-$\alpha$ DW optical depth using the expressions from \citet{Miralda-Escude_1998}, as summarized below. We consider two approaches: one based on a direct estimate from the \textsc{Thesan} simulations, and a second
one based on a useful semi-analytic approximation. In the more exact treatment, we sum up the DW optical depths across each simulation cell along random lines of sight (starting from a UV-selected galaxy) as:
\begin{align}
    \tau_{\rm DW} (\lambda_{\rm obs}) &= \frac{\tau_{\rm GP}(z_{\rm s})}{\pi} R_{\alpha} (1+\delta_{\lambda})^{3/2} \nonumber \\
    &\quad \times \sum_i x_{{\rm HI}, \, i} \left[I(x_{i}) - I(x_{i+1}) \right].
    \label{eqn:tau_dw_patchy}
\end{align}

The sum starts from the cell at the edge of the ionized bubble around the simulated UV-selected galaxy, as described below in Section \ref{subsec:bubble_sizes}. The sum then extends across all remaining simulation cells along the line-of-sight, while $x_{{\rm HI}, \, i}$ denotes the mass-weighted neutral hydrogen fraction in the $i$th cell. We map between the line-of-sight comoving coordinates of each simulation cell and redshift, moving to a new snapshot when the redshift decreases below that of the next snapshot. Until then, a sightline can wrap around a coeval snapshot according to the simulation's periodic boundary conditions. We accumulate optical depth contributions down to the redshift at which reionization completes in \textsc{Thesan} ($z = 5.5$). We note, however, that most of the DW optical depth comes from relatively nearby neutral gas, typically within a coeval snapshot.
Although the definition of the edge of the ionized bubble employed in evaluating Eqn.~\ref{eqn:tau_dw_patchy} is somewhat arbitrary, our results are insensitive to the precise definition here. This is the case provided we avoid CGM/ISM contributions from within the virial radius of each galaxy (these are considered ``intrinsic'' and impact the empirical EW distributions in our framework; we only aim to calculate the IGM optical depth contributions from first principles). 

Here $R_{\alpha}$ is a dimensionless constant related to the natural line-width of the Ly-$\alpha$ line, with $R_{\alpha} = \Lambda_{\alpha}/(4 \pi \nu_{\alpha}) = 2.02 \times 10^{-8}$, where $\Lambda_{\alpha}$ is the decay constant for the Ly-$\alpha$ line and $\nu_{\alpha}$ is the rest-frame Ly-$\alpha$ frequency, while $z_{\rm s}$ is the source redshift. The quantity $\delta_{\lambda}$ is the fractional wavelength offset defined as $\delta_{\lambda} = \Delta\lambda/\left[\lambda_{\alpha}(1+z_{\rm s}) \right]$. Hence, the observed wavelength is $\lambda_{\rm obs} = \lambda_{\alpha}(1+z_{\rm s}) + \Delta \lambda$. The Gunn-Peterson optical depth $\tau_{\rm GP}$ for fully neutral gas at the cosmic mean density is \citep{Gunn_Peterson_1965, Miralda-Escude_1998, Lidz_2021}:
\begin{align}
    \tau_{\rm GP} &= 3.8 \times 10^5 
    \left[\frac{1+z_{\rm s}}{7} \right]^{3/2} 
    \left[\frac{\Omega_b h^2}{0.0223} \right] \nonumber \\
    &\quad \times
    \left[\frac{\Omega_m h^2}{0.142} \right]^{-1/2} 
    \left[\frac{1 - Y}{0.76} \right],
    \label{eqn:tau_gp}
\end{align}
where $\Omega_m$ and $\Omega_b$ are the cosmological matter and baryon density parameters, $h$ is the Hubble parameter in the units of 100 $\mathrm{km \, s^{-1} \, Mpc}$, and $1-Y$ is the hydrogenic mass fraction. 

The DW contribution from each cell depends on $x_i = (1 + z_i)/\left[(1 + z_{\rm s})(1 + \delta_{\lambda})\right]$ and $x_{i+1} = (1 + z_{i+1})/\left[(1 + z_{\rm s})(1 + \delta_{\lambda})\right]$ where $z_i$  and $z_{i+1}$ are the starting and ending redshifts of each cell, while

\begin{equation}
    I(x_i) - I(x_{i+1}) = \int_{x_{i}}^{x_{i+1}} dx \frac{x^{9/2}}{(1 - x)^2}.
    \label{eqn:I_x}
\end{equation}

We also explore a semi-analytic description in which we determine the size of the ionized region, in a random viewing direction, surrounding each UV-emitting galaxy. We then approximate the mass-weighted neutral fraction, exterior to the ionized bubble, as spatially uniform and equal to the average value $\langle x_{\rm HI} \rangle$ from the nearest coeval \textsc{Thesan} snapshot. This amounts to taking $\tau_{\rm GP} \rightarrow \langle x_{\rm HI} \rangle \tau_{\rm GP}$, and $I(x_i) - I(x_{\rm i+1}) \rightarrow I(x_{\rm beg}) - I(x_{\rm end})$ in Eqn. \ref{eqn:tau_dw_patchy}, where
$x_{\rm beg} = (1 + z_{\rm beg})/\left[(1 + z_{\rm s})(1 + \delta_{\lambda})\right]$ is determined based on the redshift where the sightline leaves the local ionized bubble and starts to accumulate optical depth in the partly neutral IGM.
The quantity $x_{\rm end} = (1 + z_{\rm end})/\left[(1 + z_{\rm s})(1 + \delta_{\lambda})\right]$ is set by the end of reionization, approximated here as $z_{\rm end} = 5.5$.
This approach neglects spatial variations in the external neutral fraction, which is approximated by the global mean, and redshift evolution in the average neutral fraction. It includes ``patchy reionization'' effects only in accounting for the distribution of ionized bubbles around UV-luminous galaxies. Still, we find it to be a good approximation, at least for the frequencies close to line center which determine the transmission coefficients (Eqn.~\ref{eqn:transmission}). 

The key advantage of the semi-analytic method is that it allows us to explore variations around the \textsc{Thesan} reionization model. In the more exact calculation, new simulations are needed to explore
how the Ly-$\alpha$ fraction results change under different reionization models. Even then, the sensitivity to model assumptions is not always transparent. 
In the semi-analytic DW calculation, the transmission coefficients depend only on the line profiles, the bubble sizes, and the average
neutral fraction and so one can quickly explore parameter variations. 

\begin{figure*}[t]
    \includegraphics[width=\linewidth]{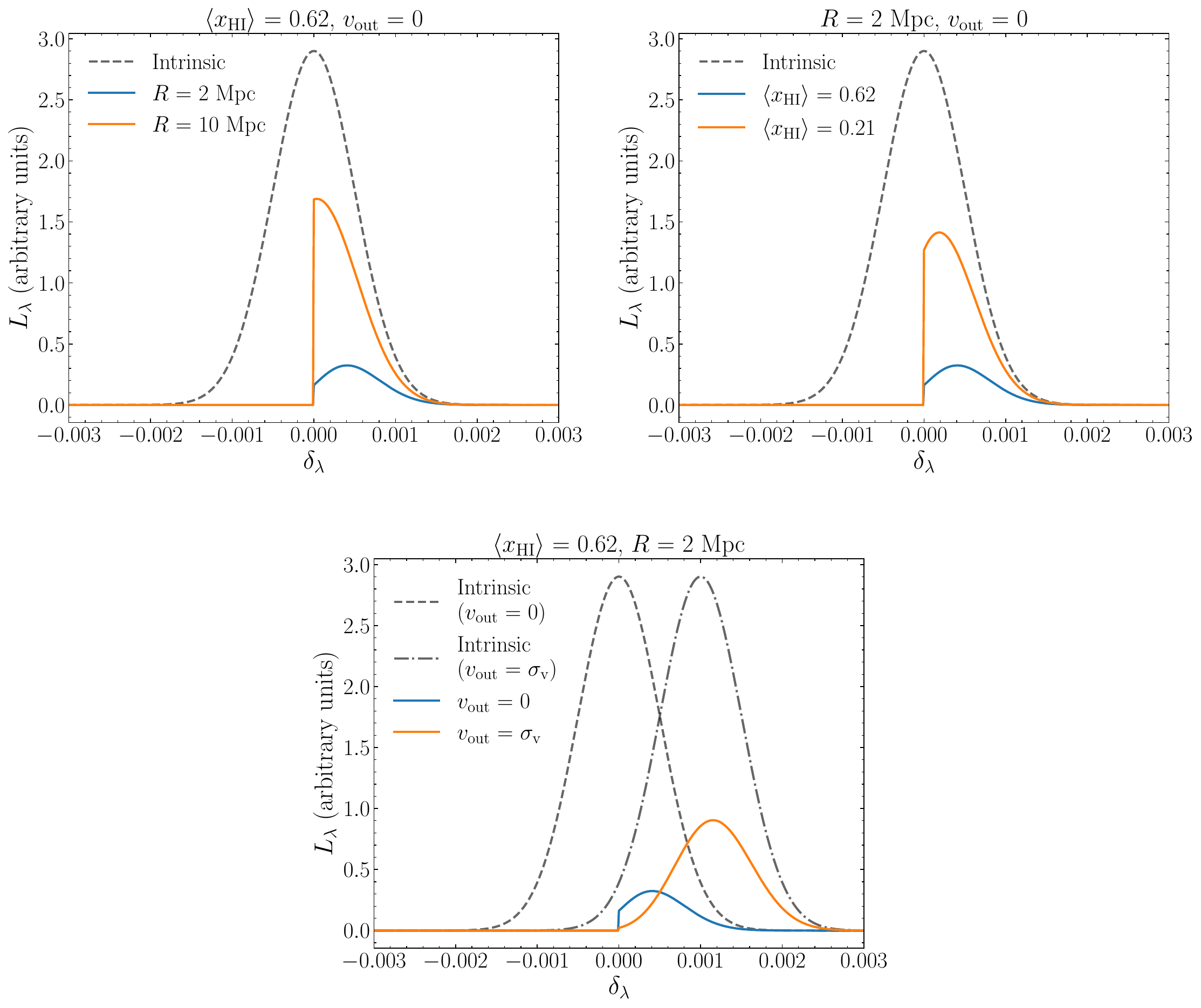}
    \caption{Dependence of the semi-analytic Ly-$\alpha$ line profiles on galaxy and IGM properties. In each panel the black dashed curves show the intrinsic line profiles without outflows. The top left panel shows trends with decreasing bubble size, with the $R=2$ comoving Mpc case (blue) showing less transmission than the $R=10$ Mpc model (orange). Here, the average mass-weighted
    neutral fraction is $\langle x_{\rm HI} \rangle = 0.62$, and there are no outflows, while $\sigma_{\rm{v}} = 150\, \rm{km\,s^{-1}}$. In each case, the region blue-ward of line center (i.e. at $\delta_\lambda < 0$) is fully absorbed due to resonant Ly-$\alpha$ absorption. 
    The top-right panel fixes $R=2$ Mpc, $v_{\rm out} = 0$, while comparing neutral fractions of $\langle x_{\rm HI} \rangle = 0.62$ (blue) 
    and $\langle x_{\rm HI} \rangle = 0.21$ (orange). The transmission shrinks in the model with a more neutral IGM due to the stronger DW absorption in this case. Finally, the bottom panel illustrates the enhanced transmission with outflows for $v_{\rm out} = \sigma_{\rm{v}} = 150\, \rm{km\,s^{-1}}$ (orange) compared to without (blue), while the dot-dashed curve is the intrinsic profile with outflows. Other model parameters are fixed as indicated in the legend. 
    }    
    \label{fig:lya_profiles}
\end{figure*}

\subsection{Measuring the Sizes of Ionized Bubbles} 
\label{subsec:bubble_sizes}

Our semi-analytic transmission calculations depend on the size of the ionized bubbles around each simulated galaxy, while the distribution of bubble sizes is also useful for understanding the results of our more exact
DW calculations. To measure the bubble sizes in \textsc{Thesan} we determine the distance traversed from
each galaxy center to the first neutral cell in a random direction. Since the bubbles are generally aspherical, this might be more accurately referred to as the length of an ``ionized skewer'' but we nevertheless use the term ``bubble size,'' while keeping in mind that our definition here is in fact a one-dimensional metric, rather than a spherically averaged quantity. 

Specifically, from each UV-luminous galaxy we draw from uniform random distributions in $\cos \theta$ and $\phi$ to specify
the spherical coordinates of a unit vector in a random direction, $\boldsymbol{\hat{n}} = (\sin \theta \cos \phi) \, \boldsymbol{\hat{x}} + (\sin \theta \sin \phi) \, \boldsymbol{\hat{y}} + (\cos \theta) \, \boldsymbol{\hat{z}}$, where $\boldsymbol{\hat{x}}$, $\boldsymbol{\hat{y}}$, and $\boldsymbol{\hat{z}}$ are unit vectors along the box axes. For each simulated galaxy, a single random line of sight is taken and the length traveled before crossing a threshold of $x_{\rm HI} > 0.5$, accounting for the simulation's periodic boundary conditions, sets the bubble size for the galaxy. The size of each simulation grid cell represents the smallest resolvable bubble. Because we apply periodic boundary conditions, a skewer can wrap around the box more than once before encountering a neutral cell near the end of reionization, and so recorded bubble sizes can exceed the box length. We determine the bubble size distributions as a function of redshift and UV magnitude.

\section{Results} 
\label{sec:results}

We now show the results of our exact DW and semi-analytic model calculations. In what follows, we refer to these loosely as the ``patchy reionization'' and ``uniform'' models, respectively, although note that both cases account for the patchy ionization structure around the simulated UV-luminous galaxies. Recall that the uniform case adopts a uniform average neutral fraction outside of the ionized skewer towards each source.

\subsection{Ly-$\alpha$ Line Profiles and Their Variations with IGM and Galaxy Properties}
\label{subsec:Lya_profile_results}

First, we show illustrative examples of the Ly-$\alpha$ line profiles after IGM scattering (see also e.g. \citealt{Santos_2004,Dijkstra_2007}). Here, we adopt the semi-analytic treatment in which the line profiles vary in a transparent way under changes in the IGM and galaxy properties. Figure~\ref{fig:lya_profiles} shows some illustrative examples, comparing the intrinsic line profile models with those after scattering in the IGM. In each case, photons blue-ward of the systemic redshift shift into line center, where they are assumed to be completely scattered.
The top left panel illustrates the importance of the size of the ionized bubbles in determining the DW optical depth. As the ionized bubble grows from $R=2$ to $R=10$ comoving Mpc (with the other model parameters fixed as indicated in the legend), the transmission increases markedly: quantitatively the peak in the specific luminosity ($L_\lambda$) grows by a factor of $\sim$5. This increase is because the Ly-$\alpha$ photons in the larger bubble redshift further into the wing of the line before encountering neutral gas. As expected, the transmission also decreases with the average neutral fraction (middle panel). 

The bottom panel illustrates how outflows can increase the transmission through the IGM. This is a consequence of the back-scattering from a neutral expanding shell in the ISM, which shifts the intrinsic line profile to the red by $\Delta \nu/\nu = -2 v_{\rm out}/c$. The shell-scattered photons reach the neutral IGM further to the red where the DW optical depth is reduced. The IGM scattering and galaxy property effects illustrated here will partly shape the Ly-$\alpha$ fraction evolution discussed below. 

\subsection{Ionized Bubble Size Distributions}
\label{subsec:bubble_size_results}

\begin{figure}[ht]
    \centering
    \includegraphics[width=\linewidth]{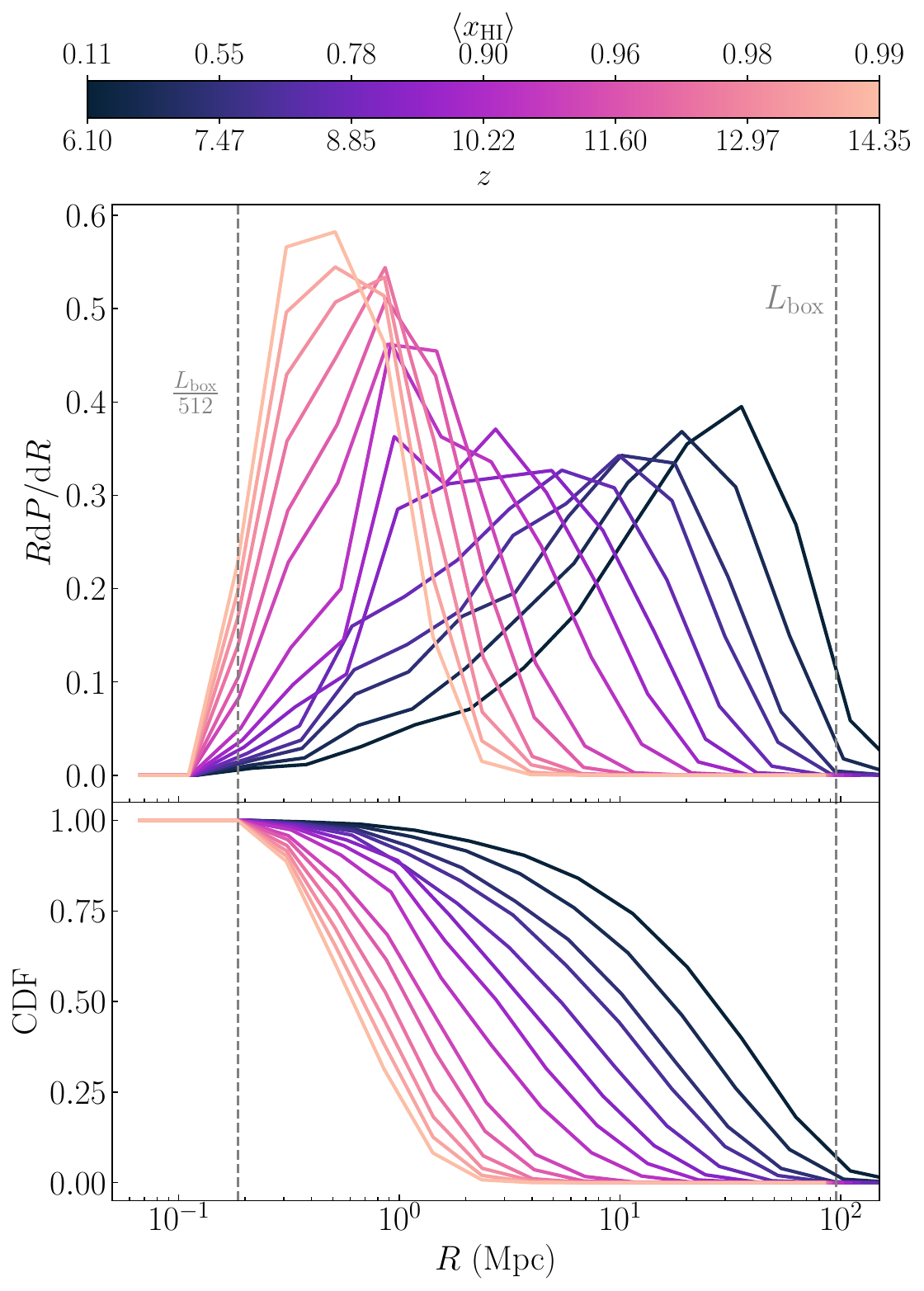}
    \caption{The redshift evolution of the ionized bubble sizes around UV-luminous galaxies in \textsc{Thesan}. The curves show the differential probability distribution function (PDF, top panel) and cumulative distribution function (CDF, bottom panel) of ionized bubble sizes, according to the definition of Section \ref{subsec:bubble_sizes}, centered on each UV-luminous simulated galaxy. The left and right vertical dashed lines indicate the cell size and box length, respectively. The color bar at the top gives the redshift and average mass-weighted ionized fraction for each curve. The bubble sizes evolve strongly as reionization proceeds, and this evolution partly shapes the Ly-$\alpha$ transmission and Ly-$\alpha$ fraction results that follow.}
    
    \label{fig:bubble_sizes}
\end{figure}

Figure~\ref{fig:bubble_sizes} shows the simulated size distribution of the ionized regions and its evolution with redshift and with the average mass-weighted neutral fraction. Here, the bubble sizes are determined using the algorithm described in Section \ref{subsec:bubble_sizes}, with random sightlines extending from each star-forming galaxy in \textsc{Thesan} that resides in a well-resolved halo with $M_{\rm halo} \geq 10^{8.5} \, M_\odot$. The characteristic bubble sizes evolve strongly with redshift. Initially, during reionization's early stages at $z \sim 13$, only small pockets of ionized gas exist with median bubble sizes of $R \lesssim 0.5$ Mpc, and high neutral fractions ($\langle x_{\rm HI} \rangle \approx 0.98$).  By $z \sim 10$ the neutral fraction reduces to about $\langle x_{\rm HI} \rangle = 0.91$ in this model, and the median bubble size is roughly $R \sim 1$ Mpc. Beyond this point, the bubbles typically contain many interior sources and the growth of the ionized bubbles is driven by the collective ionizing output from a multitude of internal sources. This process is further accelerated as neighboring large bubbles overlap, and it becomes increasingly likely for long skewers to pass entirely through highly ionized gas. For example, the median bubble size is $R \sim 5$ Mpc by $\langle x_{\rm HI} \rangle \sim 0.5$ at $z =7.4$, while it reaches $R \sim 20$ Mpc at $z=6.1$, where $\langle x_{\rm HI} \rangle = 0.1$. The \textsc{Thesan} simulation has the box length and resolution required to track this evolution, as illustrated by the gray dashed lines in the figure which show the cell size and box length of the simulations. 

\begin{figure*}[ht]
    \centering
    \includegraphics[width=\linewidth]{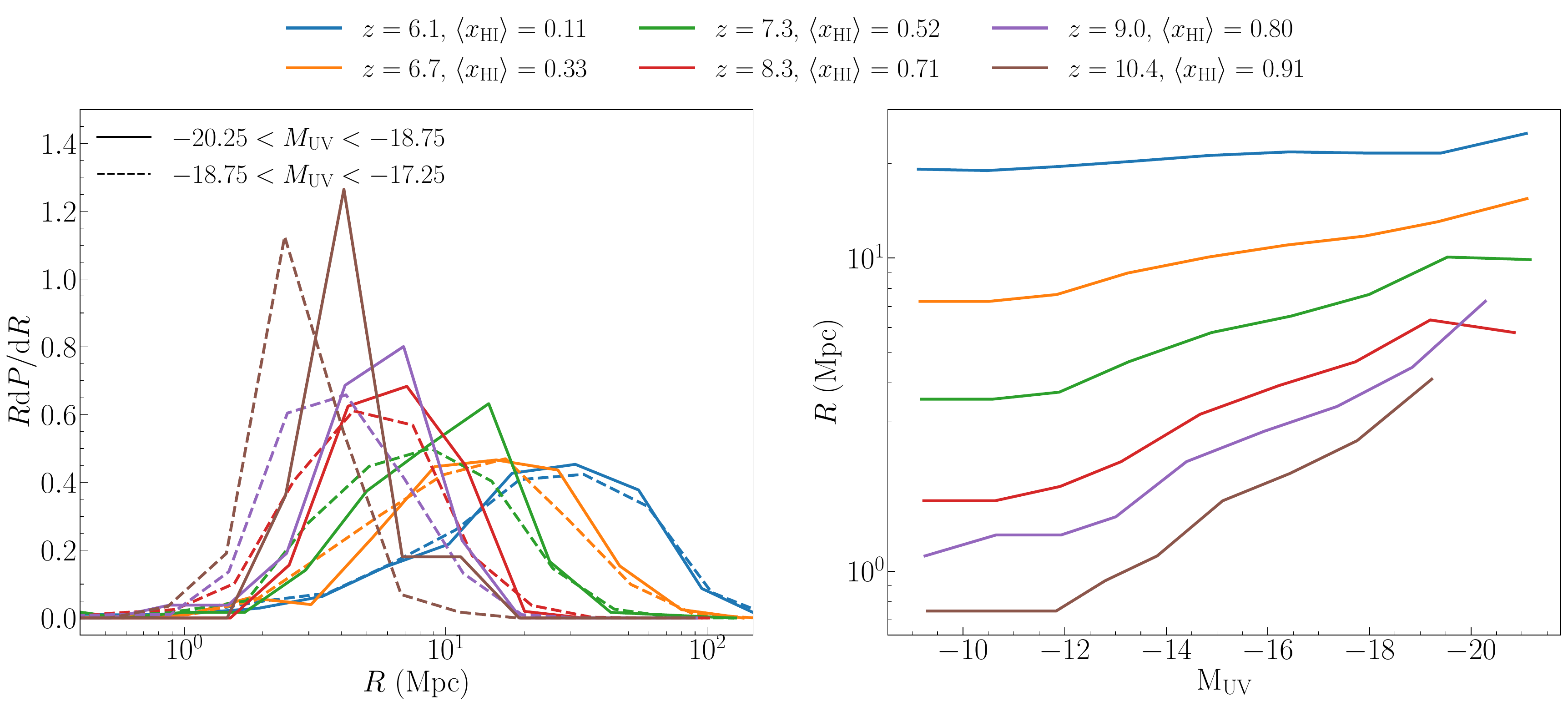}
    \caption{The size distribution of ionized bubbles in different UV luminosity bins. The left panel compares the bubble size distributions for UV magnitude bins with $-20.25 \leq M_{\rm{UV}} \leq -18.75$ (solid) and $-18.75 \leq M_{\rm{UV}} \leq -17.25$ (dashed). The right panel shows the median bubble size as a function of $M_{\rm UV}$. The redshifts and the corresponding mean neutral fractions are indicated in the legend. The brighter galaxies typically occupy slightly larger bubbles, but the trend is generally mild around the $M_{\rm UV}$ of interest for the current Ly-$\alpha$ fraction measurements (roughly those in the left panel), especially towards the end of reionization.}
    \label{fig:bubble_sizes_MUV}
\end{figure*}

Figure~\ref{fig:bubble_sizes_MUV} turns to investigate the trends of bubble size with UV magnitude. The left panel plots the bubble size
distributions in \textsc{Thesan} galaxies with $-20.25 \leq M_{\rm{UV}} \leq -18.75$ (solid) and $-18.75 \leq M_{\rm{UV}} \leq -17.25$ (dashed) at different stages of reionization. The right panel further shows the median bubble sizes as a function of $M_{\rm UV}$, across a wide range of UV magnitudes. Although the ionized bubbles generally reflect the collective influence of many sources, one still expects a statistical trend of bubble size with luminosity. More luminous sources tend to reside in massive host halos, which are highly clustered 
around prominent overdensities in the matter distribution and generally surrounded by larger ionized regions (e.g. \citealt{McQuinn_2007a}). Although the figure broadly supports these expected trends, the dependence is relatively mild across the $M_{\rm UV}$ range of interest for current Ly-$\alpha$ fraction measurements (roughly the range in the left panel). This is especially the case towards the end of reionization, where long skewers pass almost entirely through ionized gas.

\subsection{Ly-$\alpha$ Transmission Coefficients}
\label{subsec:transmission_results}

\begin{figure}[ht]
    \centering
    \includegraphics[width=\linewidth]{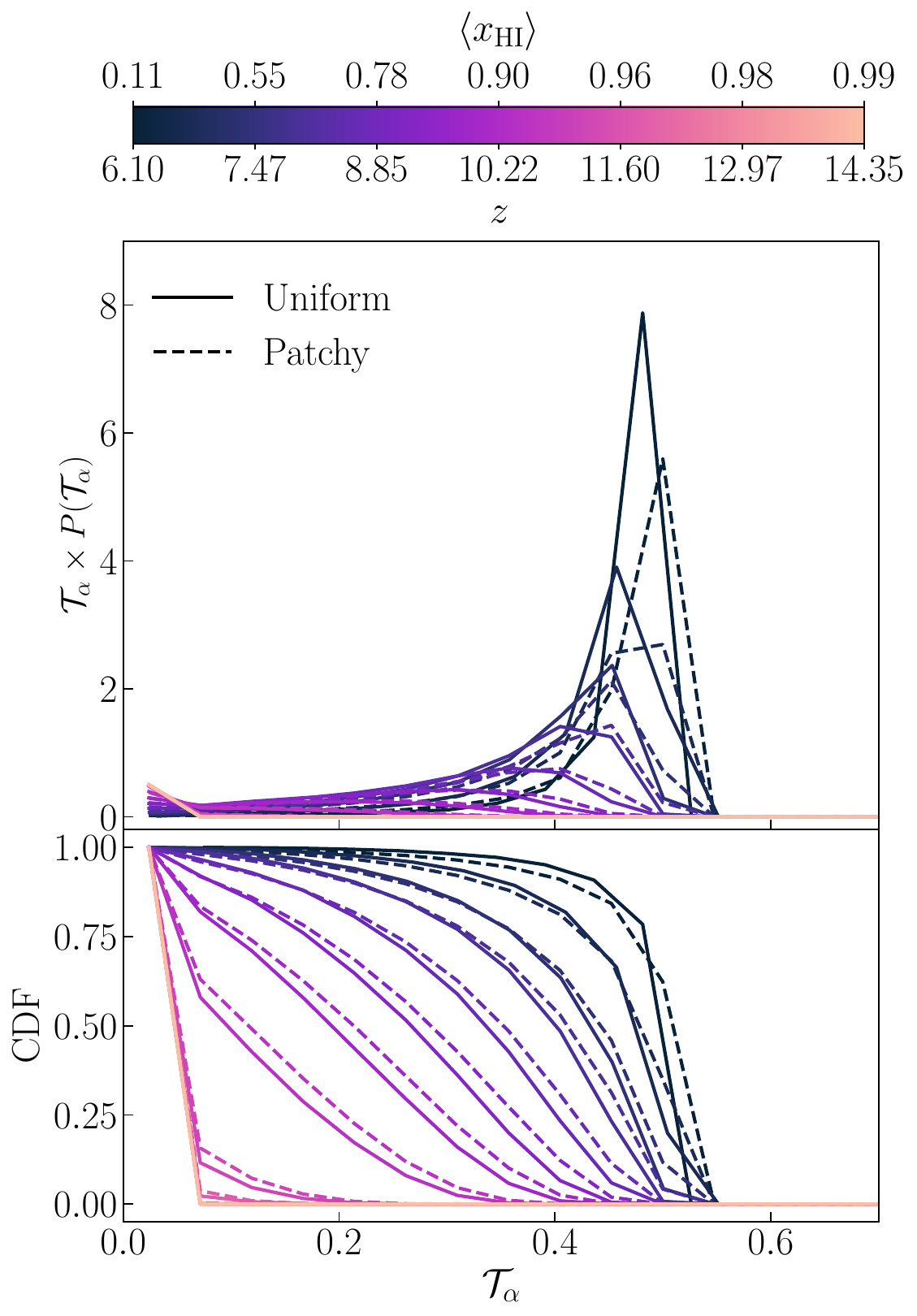}
    \caption{The distribution of Ly-$\alpha$ IGM transmission coefficients as a function of redshift and average ionization fraction. All star-forming galaxies in \textsc{Thesan} above a halo mass threshold of $M \geq 10^{8.5} M_\odot$ are included.
    The top panel shows the PDF while the bottom panel is the CDF.
    The solid curve is for the semi-analytic DW (``uniform'') calculations while the dashed curves give the exact (``patchy'') results. 
    We do not incorporate outflow velocities in the models here. 
    The transmission coefficients decrease strongly towards high redshift since the DW optical depths increase as the bubbles shrink and the neutral fraction grows.
    The transmission coefficients asymptote to $\mathcal{T}_\alpha \rightarrow 0.5$ as $\langle x_{\rm HI} \rangle \rightarrow 0$ since the blue-side of the Ly-$\alpha$ line is completely absorbed in these models. The semi-analytic and exact DW models give similar results, with the exact model showing slightly more transmission at early stages and then slightly less transmission towards the end of reionization.}
    \label{fig:transmission_dist}
\end{figure}

Figure~\ref{fig:transmission_dist} shows the PDF and CDF of the Ly-$\alpha$ IGM transmission coefficients, around all UV-selected galaxies in \textsc{Thesan} with host halo masses of $M_{\rm halo} \geq 10^{8.5} M_\odot$, as a function of redshift and average ionization fraction. Here, the models neglect outflow velocities. We contrast the semi-analytic (solid curves) and exact DW calculations (dashed curves). In each case, the transmission decreases towards high redshift, reflecting the enhanced DW optical depths from the smaller bubble sizes and larger neutral fractions in the progressively earlier stages of the EoR. In the absence of outflows, the transmission distributions become sharply peaked around $\mathcal{T}_\alpha \sim 0.5$ near the end of reionization since the blue-side of the Ly-$\alpha$ line is completely absorbed in our models. 

The figure also illustrates that the exact DW and semi-analytic predictions are quite similar, with the exact calculation giving slightly larger transmission in the early phases of reionization and slightly less transmission during the late stages. The main reason the approximate calculations provide a good description is that the transmission coefficients are only sensitive to the DW opacity close to line-center, which is largely controlled by the size of the ionized regions around each source. The size distribution of ionized regions in the semi-analytic model matches the \textsc{Thesan} results by design, modulo sensitivity to the choice of threshold neutral fraction assumed in determining the bubble sizes.
Although the exterior neutral fraction along a particular line-of-sight fluctuates around the global mean, note that in the small $\tau_{\rm DW}$ limit $e^{-\tau_{\rm{DW}}} \approx 1-\tau_{\rm DW}$, while $\tau_{\rm DW}$ is only linear in the average neutral fraction. This should also approximately apply to each sightline, provided one uses the average exterior neutral fraction along the sightline rather than the global mean. Let $\delta x_{\rm HI}$ denote the neutral fraction fluctuation around the global average along a particular sightline (exterior to the local ionized bubble). In the optically thin limit, one roughly expects the transmission fluctuation along that line of sight to scale as $\delta T_{\rm \alpha} \propto 1 - \delta x_{\rm HI}$. In this limit, the average across an ensemble of sightlines vanishes since $\langle \delta x_{\rm HI} \rangle = 0$. Therefore, the fluctuations that are neglected in the semi-analytic calculation do not influence the average transmission (to lowest order) in the optically-thin limit. The precise size distribution of neutral regions along a given sightline can also partly shape the DW absorption, but this is of relatively minor importance close to line center.  We explore the success of the semi-analytic calculation further in Appendix \ref{append_sec:sightlines_DW}. These results support using the semi-analytic calculations to explore parameter variations around the \textsc{Thesan} reionization model, at least for integrated statistics like the Ly-$\alpha$ fraction.

Although the semi-analytic calculations are useful in the present context, we caution that they are imperfect.
The optically-thin argument that fluctuations in the exterior neutral fraction average away only holds to first order. The precise distribution of neutral regions along a line of sight does partly shape the DW profiles, and replacing this with an average exterior neutral fraction does not fully capture the transmission statistics. This may be especially important for interpreting the full {\em shape} of the DW profile, rather than the integrated Ly-$\alpha$ fraction statistics considered here. We plan to consider refined semi-analytic calculations in future work. 

\begin{figure*}[t]
    \centering
    \includegraphics[width=\linewidth]{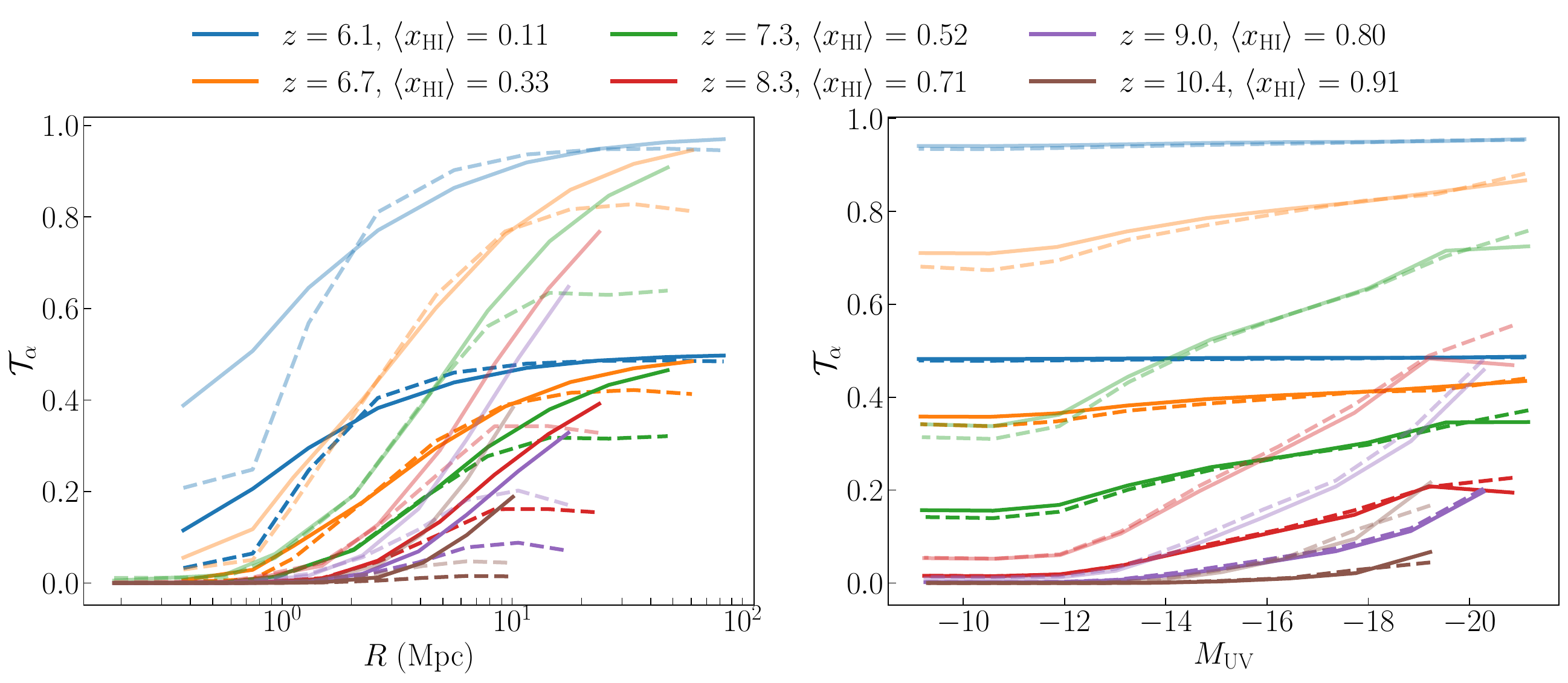}
    \caption{Median IGM transmission coefficients as a function of bubble size (left panel) and UV magnitude (right panel) for different redshifts and average
    neutral fractions (see legend). The dark shaded curves neglect outflows, while the lighter shades include outflows (with $v_{\rm out} = \sigma_{\rm v}$). The solid curves are from the semi-analytic model, while the dashed curves are the exact DW calculations.  The trends of increasing transmission with bubble size and UV magnitude are fairly well-reproduced by the semi-analytic model. Including outflows increases the transmission and this can be greater than 0.5 for large bubbles at smaller redshifts.}
    \label{fig:T_alpha_R_MUV}
\end{figure*}

Figure~\ref{fig:T_alpha_R_MUV} further shows how the median IGM transmission coefficients at several redshifts vary with bubble size (left) and UV magnitude (right). We also illustrate how outflows influence the transmission coefficients, assuming an outflow velocity of $v_{\rm out} = \sigma_{\rm v}$, with the outflow (non-outflow) cases shown in lighter (darker) shades. As expected based on our previous discussion, the median transmission coefficients increase towards low redshift as the bubbles grow larger and the neutral fractions decline. The outflows also allow enhanced transmission since the line photons mainly reach the IGM further to the red where the DW is weaker (see the bottom panel of Figure~\ref{fig:lya_profiles}). The outflow curves in the figure are hence shifted vertically upwards relative to those without outflows, with transmission coefficients greater than 0.5 for large bubbles, especially towards the end of reionization. As before, we find that the semi-analytic calculations are a good approximation to the full simulated DW results, although here one can see that the semi-analytic results tend to overestimate the transmission for small bubble sizes. See also Appendix \ref{append_sec:sightlines_DW}. 

The right panel shows qualitatively similar behavior. In this case, the increasing transmission with UV magnitude may be attributed to the
fact that brighter galaxies tend to resider in larger ionized bubbles, as discussed previously. In the context of our model, the redshift evolution of the transmission coefficients illustrated here will imprint a corresponding evolution in the observed EW distribution, as expressed through Eqn.~\ref{eq:EW_dist_cond}. The observed EW distribution and/or Ly-$\alpha$ fraction measurements can then be used to infer information about the bubble sizes, neutral fractions, and outflows. 

\subsection{Equivalent Width Distributions}
\label{subsec:EW_dist}

\begin{figure}[ht]
    \centering
    \includegraphics[width=\linewidth]{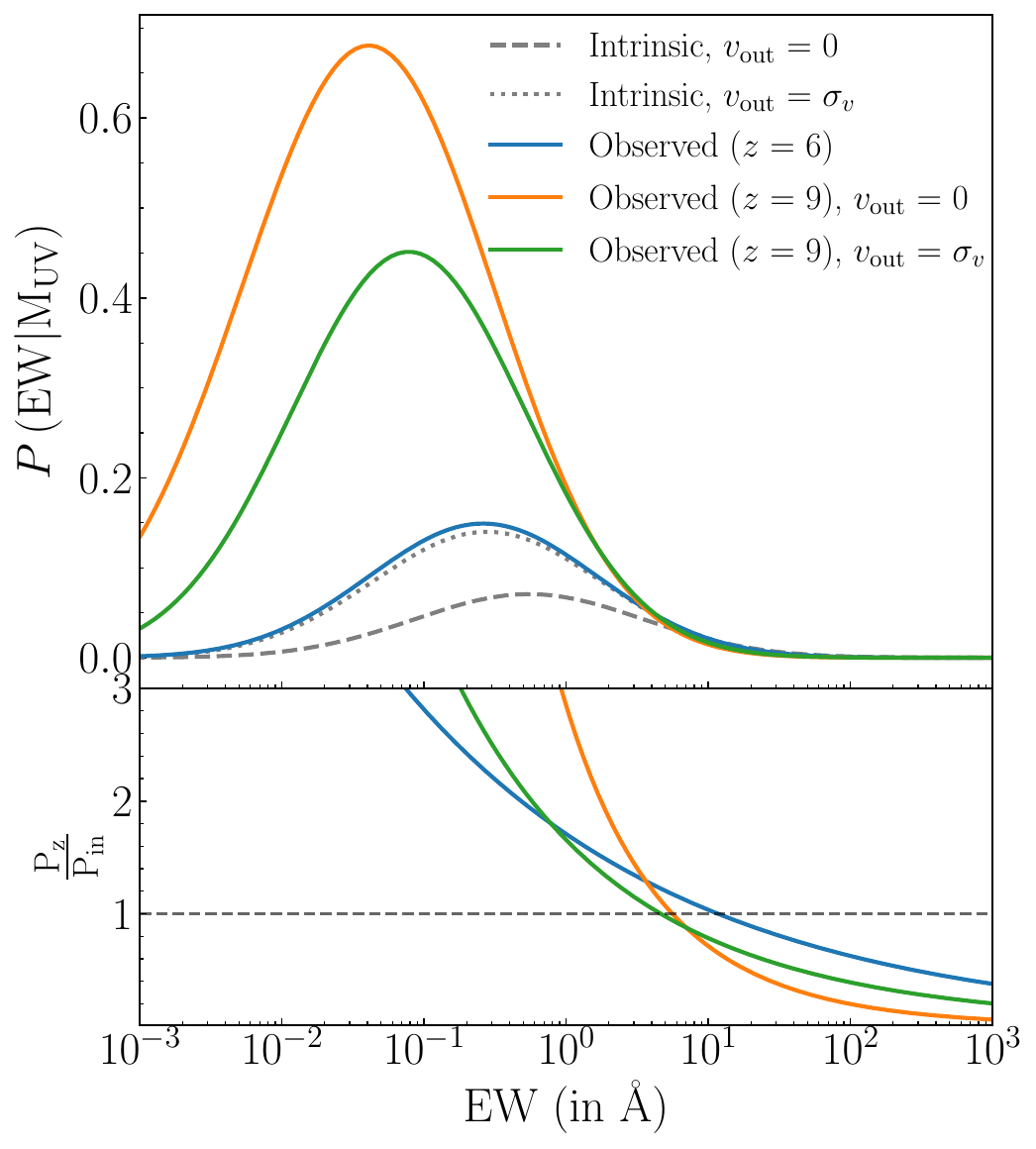}
    \caption{Observed and intrinsic EW distributions at different redshifts for galaxies with UV magnitudes between $-20.25 < M_{\rm UV} < -18.75$. In the top panel, the blue curve is the parametric fit to the observed EW distribution at $z=6$ from \citet{Tang_2024b}. The gray dotted and dashed curves show the inferred intrinsic EW distributions from our semi-analytic calculations, with (for $\sigma_v = v_{\rm circ}$) and without outflows, respectively. The green and orange curves show semi-analytic model predictions for the corresponding $\mathrm{EW}_{\mathrm{obs}}$ distributions at $z=9$. Outflows facilitate the escape of Ly-$\alpha$ photons and boost the EW values. The bottom panel shows the ratio of the observed to intrinsic EW probability distributions, while zooming-in on the larger
    EW region of the distributions which are more directly observable.}
    \label{fig:EW_dist}
\end{figure}

Figure~\ref{fig:EW_dist} shows the empirical fit to the $\mathrm{EW}_{\rm obs}$ distribution at $z=6$ for UV-selected galaxies with 
$-20.25 < M_{\rm UV} < -18.75$ from \citet{Tang_2024b},
along with our models for the intrinsic EW distribution and for the observed EW distribution at $z=9$, based on Eqns.~\ref{eq:EW_in_infer} and \ref{eq:EW_dist_cond}, respectively. The solid blue curve is the empirical fit, while the dashed gray curve shows the inferred intrinsic distribution based on our model without outflows. The dotted curve shows the intrinsic distribution for a model with an outflow velocity of $v_{\rm out} = \sigma_{\rm v}$. Since outflows increase the IGM transmission, leading to $\mathcal{T}_\alpha$ close to unity at $z=6$, the intrinsic distribution is close to the observed one for this case. In the scenario without outflows the typical transmission at $z \sim 6$ is close to $\mathcal{T}_\alpha \sim 0.5$, and so the intrinsic distribution shifts towards larger EWs by a factor of $\sim$2.
Although the distributions peak at small, unobservable EWs, they still encode key information regarding the fraction of UV-selected galaxies which lie above reasonable thresholds such as $\mathrm{EW}_{\mathrm{obs}} = 25 \, \rm{\AA}$ or $10 \, \rm{\AA}$.

The $z=9$ models show decreasing EWs due to the smaller IGM transmission coefficients at higher redshifts. Quantitatively, the median value of $\mathrm{EW}_{\mathrm{obs}}$ decreases from $\sim 8 \, \rm{\AA}$  at $z=6$ to $\sim 2.1 \, \rm{\AA}$ at $z=9$ for the no-outflows model, while the corresponding number is  $\sim 2.8 \, \rm{\AA}$ for the $v_{\rm out} = \sigma_{\rm v}$ scenario. 
Note that the impact of outflows is smaller than one might naively guess because the {\em intrinsic} EW distribution model shifts to {\em smaller} EWs with outflows, compensating partly for the enhanced $z=9$ transmission with outflows. 

\subsection{Ly-$\alpha$ Fractions}
\label{subsec:EW_fractions}

\begin{figure*}[t]
    \centering
    \includegraphics[width=\linewidth]{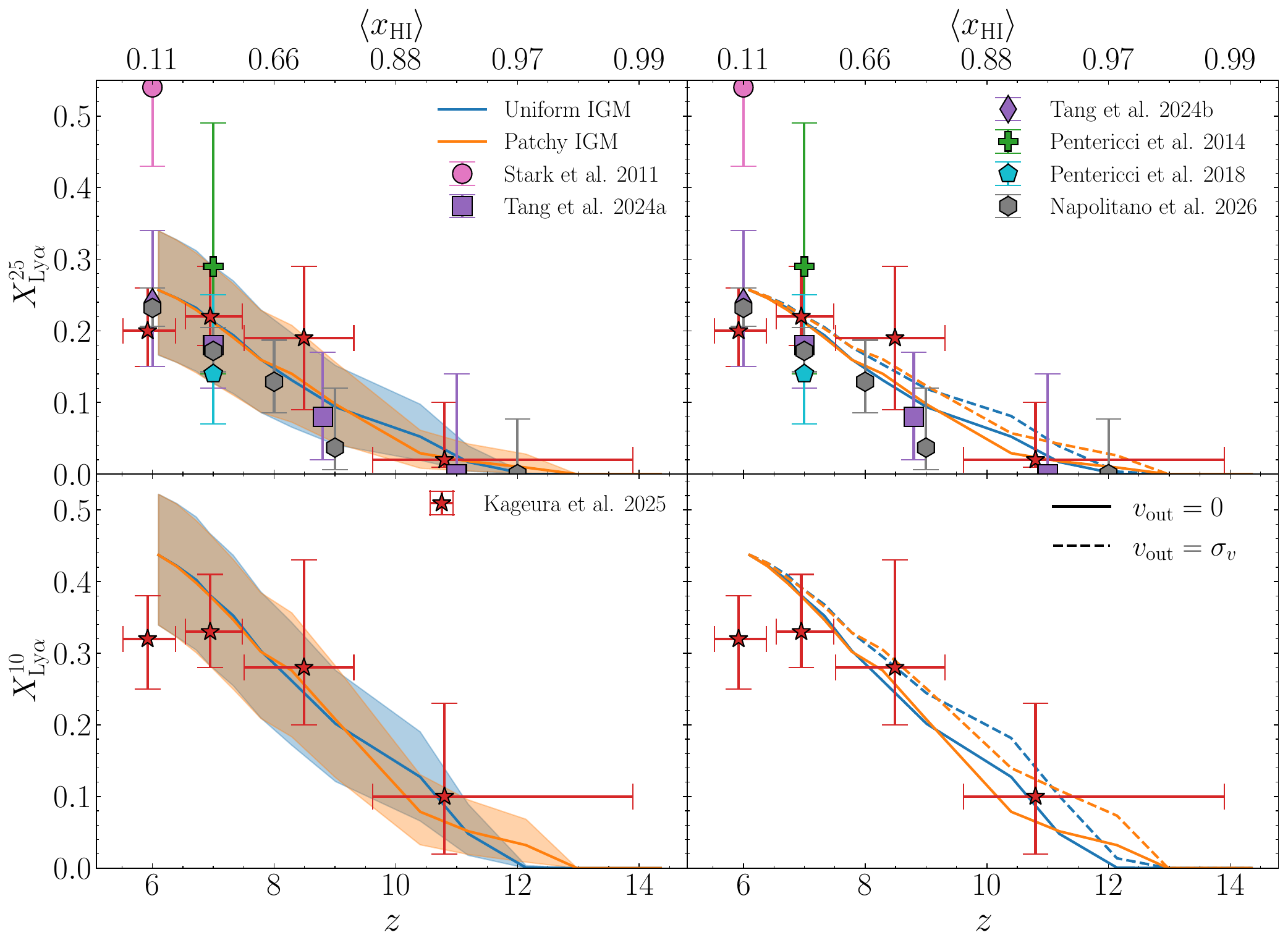}
    \caption{A comparison between Ly-$\alpha$ fraction models and observations as a function of redshift for UV-selected galaxies
    with $-20.25 < M_{\rm UV} < -18.75$.
    The top panel is for $\mathrm{EW}_{\mathrm{obs}} \geq 25 \, \rm{\AA}$, while the bottom panel takes $\mathrm{EW}_{\mathrm{obs}} \geq 10 \, \rm{\AA}$.
    The left panels show the median Ly-$\alpha$ fractions in our exact (orange) and semi-analytic (blue) models. The shaded
    regions are the 68\% confidence intervals from propagating the uncertainties on the $z=6$ empirical EW distributions.
    The right panel contrasts the median results in models with (dashed) and without (solid) outflows. The points, with $1\sigma$ uncertainties, are observational estimates from \citet{Kageura_2025}, \citet{Tang_2024a}, \citet{Tang_2024b}, \citet{Pentericci_2014}, \citet{Pentericci_2018}, and \citet{Napolitano_2026b}. The average mass-weighted neutral fractions from \textsc{Thesan} are shown along the upper x-axis for reference. Although the measurement uncertainties remain large, our models are consistent with the observational data, supporting the reionization history and bubble size distribution models in \textsc{Thesan}.}
    \label{fig:X_Lya_fiducial}
\end{figure*}

Figure~\ref{fig:X_Lya_fiducial} shows the resulting Ly-$\alpha$ fraction model predictions, as a function of redshift and neutral fraction for UV-selected galaxies with $-20.25 < M_{\rm UV} < -18.75$, compared to current observational estimates. The Ly-$\alpha$ fraction depends on the $\mathrm{EW}_{\rm obs}$ threshold, $\mathrm{EW_{th}}$, imposed for a galaxy to be counted as a Ly-$\alpha$ emitter (LAE), and so the Ly-$\alpha$ fraction is denoted  $X_{\rm Ly\alpha}^{\rm EW_{th}}$ and we consider $\rm{EW_{th}} = \rm{25 \, \AA}$ (top panel) and $\rm{EW_{th}} = \rm{10 \, \AA}$ (bottom panel). 
The solid blue and orange curves show the median model predictions from our semi-analytic (``uniform") and exact IGM (``patchy") DW calculations, respectively. The shaded regions indicate the $68\%$ confidence interval uncertainties in the model predictions based on propagating the error bars from the $z=6$ measurements of \citet{Tang_2024b}. Therefore, the band only accounts for the uncertainties in the EW distributions that our models are anchored to, while sample variance and Poisson errors are assumed to be properly included in the reported Ly-$\alpha$ fraction measurement uncertainties.\footnote{One caveat is that the sample/cosmic variance may be large given the relatively small JWST fields and may not be fully accounted for in the observational error budgets. See \citet{Taylor_2014} for relevant discussions.}
The model predictions are compared
to current observational estimates from \citet{Kageura_2025}, \citet{Tang_2024a}, \citet{Tang_2024b}, \citet{Pentericci_2014}, \citet{Pentericci_2018}, and \citet{Napolitano_2026b}.

The main message behind this figure is that our models match the observational estimates well across the full redshift range covered from
$z \sim 6-14$, with $\chi_\nu^2 \approx 0.4-1.2$, depending on the EW threshold. Although the models are calibrated to the $z = 6$ EW distributions from \citet{Tang_2024b}, and so match their $z=6$ Ly-$\alpha$ fraction estimates by construction, the agreement at higher redshifts is non-trivial. The figure further shows that the exact and semi-analytic DW model predictions are quite similar, as expected based on the agreement between the transmission distributions in these two cases (see Section \ref{subsec:transmission_results} and Appendix \ref{append_sec:sightlines_DW}). The right panel of Figure~\ref{fig:X_Lya_fiducial} also illustrates how outflows only slightly increase the modeled Ly-$\alpha$ fractions. This insensitivity to outflows is because of the compensating effect that outflows have on our intrinsic EW distribution calibration, as discussed in Section \ref{subsec:EW_dist}. 

\begin{figure}[t]
    \centering
    \includegraphics[width=\linewidth]{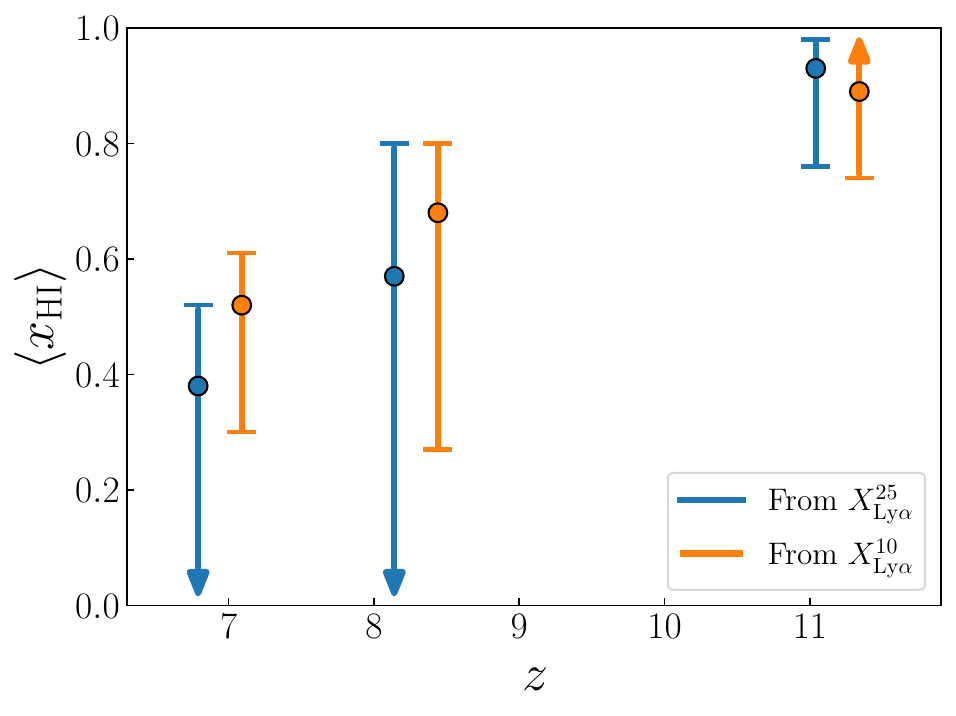}
    \caption{The range of allowed average mass-weighted neutral fractions in our fiducial semi-analytical models derived using the JWST measurements from \citet{Kageura_2025}. The blue and orange points are from $X^{25}_{\rm{Ly\alpha}}$ and $X^{10}_{\rm{Ly\alpha}}$ measurements, respectively. These results are in good agreement with the estimates of \citet{Kageura_2025} and \citet{Mason_2026}.}
    \label{fig:constraints}
\end{figure}

Given the success of the semi-analytic calculations, we use this approach to derive constraints on the neutral hydrogen fraction using
the Ly-$\alpha$ fraction measurements from \citet{Kageura_2025} at $z \sim 7, 8$ and $11$\footnote{There are more recent JWST measurements by \citet{Napolitano_2026b}, which agree well with \citet{Kageura_2025} but only include $\mathrm{EW_{th}}$ $>25\,\rm{\AA}$.}. Specifically, we vary the model average mass-weighted neutral fractions, while fixing the bubble size distributions to the \textsc{Thesan} values at matching neutral fractions (yet generally differing redshifts). That is, we are allowing shifts in the redshift at which a given neutral fraction is reached, while assuming the reionization morphology/bubble sizes are fixed by the average neutral fraction. 
Figure~\ref{fig:constraints} shows the results of independently fitting to the \citet{Kageura_2025} measurements for each of
$\mathrm{EW_{th}}$ $>25\,\rm{\AA}$ (blue) and $>10\,\rm{\AA}$ (orange). The best-fit and $1\sigma$ average neutral bounds from the
$X_{\rm{Ly\alpha}}^{25}$ measurements are: $\langle x_{\rm{HI}} \rangle$ = $0.38^{+0.14}_{-0.38}$, $0.57^{+0.23}_{-0.57}$, and $0.93^{+0.05}_{-0.17}$ at $z \sim$ 7, 8, and 11, respectively. The respective values derived from the $X_{\rm{Ly\alpha}}^{10}$ measurements are: $0.52^{+0.09}_{-0.21}$, $0.68^{+0.12}_{-0.39}$, and $0.89^{+0.11}_{-0.15}$. Note that we do not attempt to derive bounds from the Ly-$\alpha$ fraction measurements at $z \sim 6$ since our EW distribution models are anchored at this redshift. 

Although the neutral fraction uncertainties are still large, the Ly-$\alpha$ fraction measurements are starting to provide
valuable information regarding the reionization history, with good prospects for further improvements. 

\subsection{Sensitivity to the EW Distribution Model}
\label{sec:ew_dist}

Our fiducial model adopts the $z = 6$ log-normal EW distribution with parameters from \citet{Tang_2024b}. Here, we briefly test the impact of instead using alternate empirical EW distribution models from the current literature. Specifically, we use an exponential distribution with parameters taken from \citet{Mason_2018}. Within the magnitude bin $-20.25 < M_{\rm UV} < -18.75$, we find the log-normal distribution model to be more skewed towards smaller EWs. On the other hand, the exponential distribution has higher probability between $20\, \rm{\AA} \lesssim \rm{EW} \lesssim 200\, \rm{\AA}$. Although the log-normal distribution has a heavier tail at larger EWs, the tail probabilities are small enough that this difference has little impact on the Ly-$\alpha$ fraction predictions. In the \citet{Mason_2018} model, we find higher Ly-$\alpha$ fractions at $z \lesssim 8$ by $\approx 0.1-0.2$, but the results still fall within the observed uncertainties. 

\subsection{Comparison with Previous Work}
\label{sec:comp_other_work}

It is instructive to compare our results with those from \citet{Neyer_2025}, who also calculate Ly-$\alpha$ fractions during reionization from the same \textsc{Thesan} simulation. Their work also shows consistency between the simulated Ly-$\alpha$ fractions at $z \gtrsim 8$ and observations for an EW threshold of $\rm{EW}_{\rm obs} \geq 25 \, \rm{\AA}$. In contrast to our results, however, those authors find larger Ly-$\alpha$ fractions at $z \sim 6-8$ than in most previous observations, although their results agree better with the early $z = 6$ estimates from \citet{Stark_2011}. In their $M_{\rm UV} = -19$ bin, the $\rm{EW} \geq 10 \, \rm{\AA}$ results also exceed the observational ones from \citet{Kageura_2025} at all redshifts considered.

We attribute the better agreement between models and observations found here to differences in our approaches for calibrating Ly-$\alpha$ photon escape fractions. As in our work, an empirical step is needed for \citet{Neyer_2025} to account for dust attenuation and Ly-$\alpha$ photons that are scattered out of the line-of-sight by neutral hydrogen on poorly-resolved ISM and CGM scales. Their work employs a multi-parameter model for outflows and dust attenuation, which is calibrated using measurements of the $z=5.5$ and $z=6.6$ Ly-$\alpha$ emitter luminosity functions from narrow-band LAE surveys, carried out using the Hyper-Supreme Cam on the Subaru telescope \citep{Ouchi_2008,Ouchi_2010,Konno_2018}. A limitation of this approach is that the LAE luminosity function measurements only span a limited range in Ly-$\alpha$ luminosities at the bright-end, with $L_\alpha \sim 10^{42.5}-10^{43.8}$ erg/s. That is, the Ly-$\alpha$ fractions are partly determined by UV-selected galaxies which emit weak Ly-$\alpha$ lines, while their model anchors only to observations of relatively bright LAEs.

Quantitatively, we can consider which galaxies are captured in their calibration to $L_\alpha \gtrsim 10^{42.5}$ erg/s LAEs. The observed UV magnitude range of UV-selected galaxies in \citet{Kageura_2025} is $-20.25 \leq M_{\rm UV} \leq -18.75$. Assuming a UV continuum luminosity with $L_\lambda \propto L^{-\beta -2}$ and $\beta=-2$, only galaxies with $\mathrm{EW}_{\rm obs} \geq 113 \, \rm{\AA}$ exceed the minimum Ly-$\alpha$ luminosity threshold of their calibration sample at $M_{\rm UV} = -18.75$, while $\mathrm{EW}_{\rm obs} \geq 28 \,  \rm{\AA}$ are captured for $M_{\rm UV} = -20.25$ (see Eqn.~\ref{eq:ew_def}). These numbers can be compared to the median observed EWs determined empirically by \citet{Tang_2024b}. In observed bins with median UV luminosities of $M_{\rm UV} = -19.5$, ($-18.5$), the median observed EWs at $z=5-6$ are $10 \pm{2} \, \rm{\AA}$ ($16 \pm 3 \, \rm{\AA}$). Thus, their calibration procedure effectively anchors to the high EW tails of the distribution, beyond the observed median Ly-$\alpha$ line strengths.
A concern, then, is that the calibration requires significant extrapolations to describe the more typical Ly-$\alpha$ line strengths among UV emitters. Moreover, although the qualitative trend in their models of increasing Ly-$\alpha$ line strength with decreasing UV luminosity is seen empirically \citep{Tang_2024b, Qin_2025}, their model overproduces the median $\mathrm{EW}_{\rm obs}$, especially among fainter galaxies. For example, for an observed $M_{\rm UV} = -18.5$, their Figure 3 suggests a median observed EW at $z=6$ of $\mathrm{EW}_{\rm obs} \sim 70 \, \rm{\AA}$, considerably larger than the \citet{Tang_2024b} estimate of $16 \pm 3 \, \rm{\AA}$ for this UV luminosity and $z \sim 5-6$.

In contrast, we explicitly anchor to the EW distribution at $z = 6$. We therefore reproduce the $z \sim 6$ Ly-$\alpha$ fraction observations by construction. Importantly, our results also match the full observed redshift evolution at higher redshifts, supporting the reionization history and bubble size distributions in \textsc{Thesan}, under the assumption that the intrinsic EW distribution is redshift invariant. In the absence of challenging first-principles predictions for the Ly-$\alpha$ escape fractions, our approach of anchoring to empirical estimates of the Ly-$\alpha$ EW distribution, and predicting the evolution into the EoR, seems more reliable than calibrating to the bright-end of the $z \sim 6$ Ly-$\alpha$ luminosity function. 

We also repeated our calculations using the intrinsic Ly-$\alpha$ profile from \citet{Neyer_2025}, which is double-peaked and also accounts for outflows. The peak red-ward of line center in this alternative profile lies between the peaks of our Gaussian profiles with $v_{\rm{out}} = 0$ and $\sigma_v$, respectively. Adopting the \citet{Neyer_2025} line profile thus leads to similar Ly-$\alpha$ fraction results, intermediate between our results with/without outflows. This supports the conclusion that the differences in our intrinsic line luminosity calibration procedures are primarily responsible for the differences between our Ly-$\alpha$ fraction results. 

\citet{Kageura_2025} also derive constraints on the neutral hydrogen fraction using a similar approach to that used here. Those authors calibrate an intrinsic Ly-$\alpha$ line emission model based on their own $z \sim 5$ EW distribution measurements, and use \textsf{21cmFAST} semi-numerical simulation outputs to model Ly-$\alpha$ scattering in the IGM at $z \gtrsim 6$. Despite the difference in our methodologies, our best-fit $\langle x_{\rm HI} \rangle$ values at $z \sim$ 7, 8, and 11 agree with their estimates within $0.8\sigma$ ($X^{25}_{\rm{Ly\alpha}}$) and $0.4\sigma$ ($X^{10}_{\rm{Ly\alpha}}$), and our $1\sigma$ confidence intervals are comparable. Our results also agree with those of \citet{Mason_2026}, who likewise use semi-numerical simulations to interpret Ly$\alpha$ fraction measurements, within $0.3\sigma$ ($X^{25}_{\rm Ly\alpha}$) and $0.7\sigma$ ($X^{10}_{\rm Ly\alpha}$). The agreement across these independent analyses is encouraging. 

\section{Discussion and Conclusions}
\label{sec:conclusions}

We have explored the implications of recent JWST measurements of the fraction of UV-selected galaxies which emit Ly-$\alpha$ lines at $z \sim 6-14$. In order to determine the Ly-$\alpha$ fraction evolution into the EoR, our model anchors to empirical estimates of the Ly-$\alpha$ EW distribution near the end of reionization from \citet{Tang_2024b}. We adopt a Gaussian line profile model, allowing for velocity offsets from outflows, to describe the Ly-$\alpha$ spectral shape for photons leaving the ISM and CGM of the host galaxy. We subsequently model the IGM scattering using the \textsc{Thesan} simulations, and predict the observed EW distributions under the assumption that the intrinsic (i.e. before IGM scattering) EW distribution is redshift independent from $z = 6-14$. Our results are summarized as follows:

\begin{itemize}

\item After calibrating to the observed EW distribution at $z=6$, our model predictions are in good agreement with the observed Ly-$\alpha$ fraction evolution from $z=6-14$, with $\chi^2_\nu = 0.4\,(1.2)$ for the $X_{\rm{Ly\alpha}}^{25}$ ($X_{\rm{Ly\alpha}}^{10}$) measurements from \citet{Kageura_2025}. This broadly supports the reionization history and bubble size models in the \textsc{Thesan-1} simulation (see Figure~\ref{fig:bubble_sizes}). 

\item Although outflows increase the IGM transmission, their impact on the Ly-$\alpha$ fraction model predictions is small after we calibrate to the observed $z=6$ EW distribution. In models with outflows, the inferred intrinsic EW distribution shifts to smaller values, partly compensating for the enhanced IGM transmission.

\item We find that a semi-analytic description of the IGM transmission coefficients mostly reproduces our more exact DW calculations from \textsc{Thesan}. The semi-analytic calculations use the bubble size distributions from \textsc{Thesan}, while they approximate the exterior IGM mass-weighted neutral fractions by the global average values. On the other hand, the exact DW calculations account for the precise distribution of neutral hydrogen along each line-of-sight.  Although the semi-analytic approximation is adequate for interpreting current Ly-$\alpha$ fraction measurements, refinements may be required in the future, especially for observables sensitive to the precise shape of the DW profile.

\item After extracting the bubble size distributions from \textsc{Thesan}, we use the semi-analytic calculations to explore variations around the reionization history in \textsc{Thesan}. 
The $1\sigma$ average mass-weighted neutral fraction constraints derived from the $X_{\rm{Ly\alpha}}^{25}$ measurements from \citet{Kageura_2025} are: $\langle x_{\rm{HI}} \rangle$ = $0.38^{+0.14}_{-0.38}$, $0.57^{+0.23}_{-0.57}$, and $0.93^{+0.05}_{-0.17}$ at $z \sim$ 7, 8, and 11, respectively. The corresponding values derived from the $X_{\rm{Ly\alpha}}^{10}$ measurements are: $0.52^{+0.09}_{-0.21}$, $0.68^{+0.12}_{-0.39}$, and $0.89^{+0.11}_{-0.15}$. 

\item At $z \gtrsim 8$ our results are mainly consistent with the earlier study of \citet{Neyer_2025} from the same simulation. At $z \sim 6-8$, however, we find better agreement with observations than those authors. We attribute this difference to our calibrating directly to the observed $z \sim 6$ EW distributions, while the earlier work only explicitly matches to the bright end of the LAE luminosity function. Our results agree with those from \citet{Kageura_2025} and \citet{Mason_2026} even though our models employ different reionization simulations and EW distributions. 

\end{itemize}

We emphasize that we assume the intrinsic EW distribution to be redshift-independent between $z \sim 6$ and $z \sim 14$. However, this distribution may evolve due to a declining dust abundance towards higher redshifts \citep{Peroux_2020,Cullen_2024}, which would boost the typical intrinsic EW and require a more rapid increase in the IGM neutral fraction towards high redshift in order to match the Ly-$\alpha$ fraction observations. On the other hand, there is growing evidence for large neutral hydrogen columns in the ISM/CGM of some early galaxy populations \citep{Heintz_2024,Heintz_2025}, which may suppress Ly-$\alpha$ escape at high redshift and cause our model to overestimate the IGM neutral fractions required to match observations. Future modeling efforts, along with empirical efforts to determine dust abundances and ISM/CGM neutral hydrogen column densities, will be required to sharpen inferences of the  IGM neutral fractions during reionization. 

There are outstanding observational prospects for improving the Ly-$\alpha$ fraction measurements in the near future, and for carrying out complementary analyses of Ly-$\alpha$ emitting galaxy populations.  
First, additional JWST observations will help in assembling larger galaxy samples, especially during the early stages of the EoR at $z \gtrsim 10$ where current measurements probe only a small handful of galaxies.
Second, the Prime Focus Spectrograph on the Subaru telescope recently saw its first light and will deliver unprecedented LAE samples into the EoR \citep{Greene_2022}.
Third, Roman space telescope Ly-$\alpha$ grism spectroscopy will enable measurements over much wider regions on the sky \citep{Wold_2024} mitigating cosmic/sample variance. We also anticipate future improvements in measurements of  
the LAE luminosity function evolution (e.g. \citealt{Ota_2017}), LAE spatial clustering \citep{Furlanetto_2006b,McQuinn_2007b}, the LAE void distribution \citep{Gangolli_2021}, 
cross-correlations with 21 cm surveys \citep{Lidz_2009,LaPlante_2023} and correlations with line-intensity maps \citep{Cheng_2022},  
among other analyses.  Taken together, these new data sets and analyses will provide further powerful tracers of the reionization history and the spatial structure of the reionization process.

\begin{acknowledgments}
AK and AL acknowledge support from NASA grant 80NSSC26K0183. GS acknowledges support from a CIERA Postdoctoral Fellowship, with additional support provided by NSF through grant AST-2307327; by NASA through grant 23-ATP23-0008; and by STScI through grant JWST-AR-03252.001-A. We thank Josh Borrow for initial help with reading the \textsc{Thesan} simulation outputs.

\section*{Use of Large Language Model Acknowledgment}

We acknowledge the use of Claude Sonnet 4.6 and Opus 4.8 in optimizing the codes employed in this work. In particular, we used these LLMs to convert multiple different codes into a coherent Python package\footnote{\url{https://github.com/aritrak-98/Lyman_alpha_DW_modeling.git}} that is publicly available for use and will be continuously updated. We vetted the returned outputs before making it available to the community.
\end{acknowledgments}

%

\software{
astropy \citep{Astropy_2013, Astropy_2018, Astropy_2022},  
matplotlib \citep{Matplotlib_2007},
IPython \citep{IPython_2007},
NumPy \citep{Numpy_2011, Numpy_2020}, 
SciPy \citep{Scipy_2020},
NASA's Astrophysics Data System (ADS),
arXiv
}


\appendix

\restartappendixnumbering

\section{Exact and Semi-Analytic Damping Wing Optical Depth Profiles}
\label{append_sec:sightlines_DW}

\begin{figure*}[ht!]
    \centering
    \includegraphics[width=0.9 \linewidth]{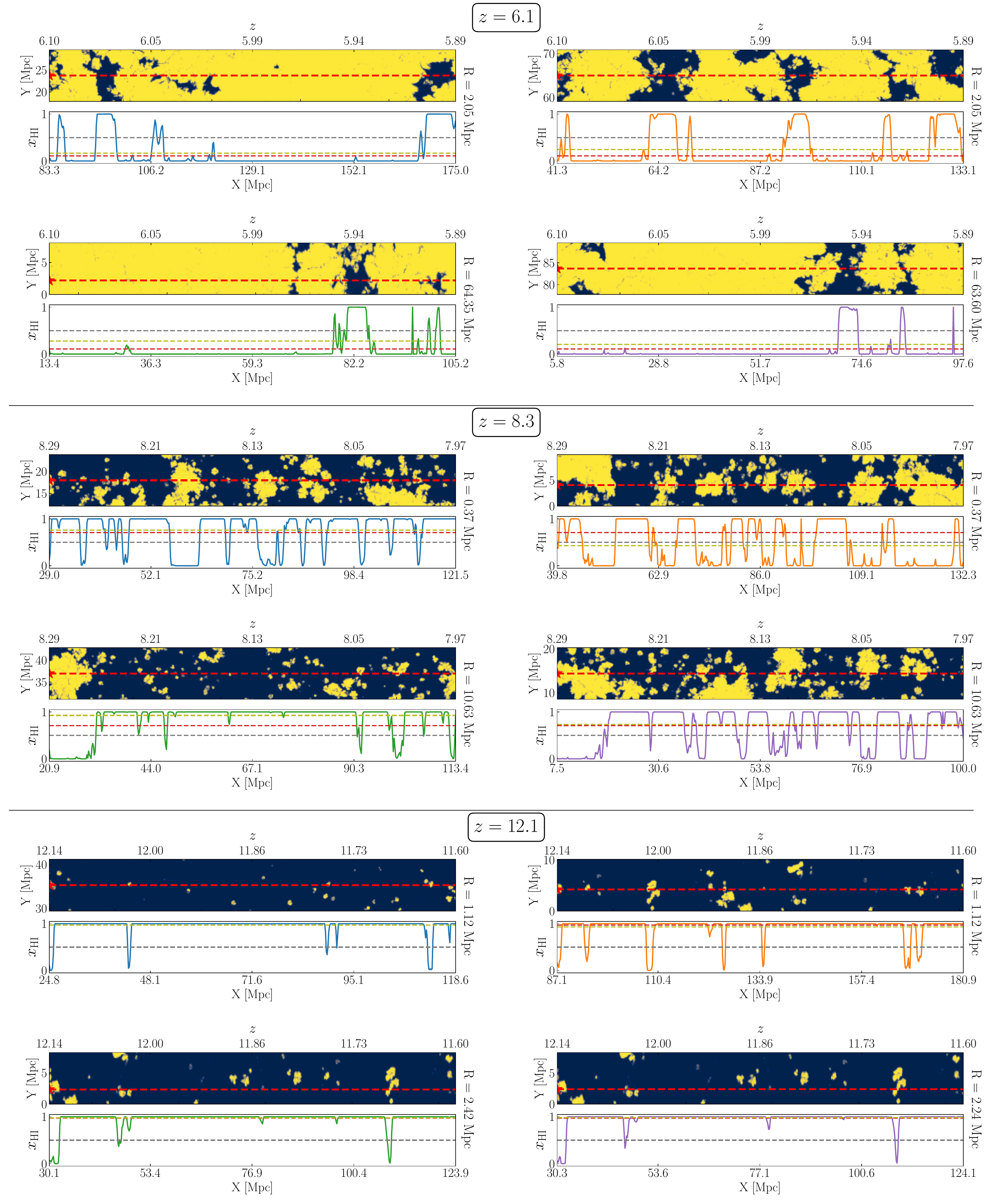}
    \caption{Four example sightlines at each of $z = 6.1$, $z=8.3$, and $z=12.1$ towards UV-selected galaxies in the simulations. In each sub-panel, the top rows show narrow slices through the simulation data cubes with neutral hydrogen in dark blue and ionized regions in yellow. The bottom rows show the neutral hydrogen fraction ($x_{\rm HI}$) along each line of sight. We pair sightlines with matching bubble sizes to isolate the impact of exterior neutral fraction variations.  
    The horizontal gray line shows the threshold neutral fraction used to define the size of the ionized region around each galaxy in the semi-analytic model ($x_{\rm HI} = 0.5$), while the red-dashed line is the average neutral fraction $\langle x_{\rm HI} \rangle$ of the coeval box at the corresponding redshift, adopted in the semi-analytic calculation. The olive-dashed lines show estimates of the average neutral fraction, exterior to the first ionized region, along each line-of-sight. This illustrates the sightline-to-sightline scatter around the global average neutral fraction. 
    The bubble sizes for each sightline are labeled on the far right of the sub-panels. 
    The sightlines illustrate the diversity of bubble sizes and of exterior neutral fractions in the simulations.
    }
    \label{append_fig:los_examples}
\end{figure*}

\begin{figure*}[ht!]
    \centering
    \includegraphics[width=\linewidth]{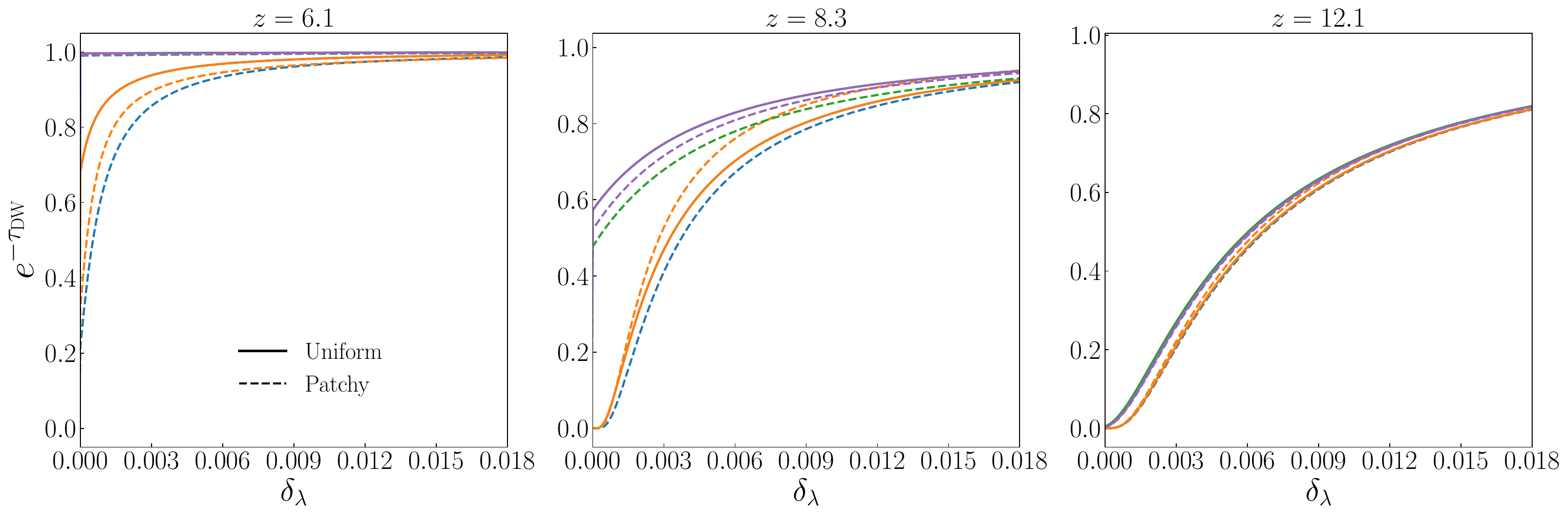}
    \caption{
    DW transmission curves, as a function of wavelength offset, for the same sightlines shown in the previous figure. The color-code matches that of the sightlines in the previous figure and so these curves indicate the DW profiles for the corresponding sightlines.  
    The dashed curves give the exact DW calculations (Eqn.~\ref{eqn:tau_dw_patchy}), while the solid curves are semi-analytic model results. The semi-analytic models are specified by $\langle x_{\rm HI} \rangle$ and $R$, so some of the solid curves (labeled ``uniform'') overlap.
    Note that the transmission coefficients generally sample only $\delta_\lambda \lesssim 0.002$, but we show a wider range of wavelength offsets for completeness. Although the DW curves at a given redshift are diverse, the spread in the transmission close to line center is mainly explained by the variations in bubble size. This helps to understand the success of the semi-analytic description.}
    \label{append_fig:dw_examples}
\end{figure*}

In the body of the text, we found fairly good agreement between the exact simulated DW transmission coefficients (calculated using Eqn.~\ref{eqn:tau_dw_patchy}) and a semi-analytic model. In order to gain some intuition for {\em why} the semi-analytic description is successful, it is helpful to examine example sightlines through the simulated neutral hydrogen distribution and to inspect the corresponding exact DW transmission profiles, along with those in the semi-analytic approximation (see Figures \ref{append_fig:los_examples} and \ref{append_fig:dw_examples}). 

Specifically, we consider four example sightlines drawn from each of $z=6.1$, $z=8.3$, and $z=12.1$ towards UV-selected galaxies, where the average mass-weighted neutral fractions are $\langle x_{\rm{HI}} \rangle$ $= 0.11$, $0.71$, and $0.97$, respectively. At each redshift, we consider two pairs of sightlines with matching bubble sizes (bubble size matches are shown along rows). As in the body of this paper, by ``bubble size'' here we mean the length traversed from the galaxy to the first neutral region along the line-of-sight, based on when the neutral fraction first crosses a threshold value of $x_{\rm HI} \geq 0.5$. Hence, comparing the examples from left to right along a row reveals the spread in exterior neutral fractions beyond the local ionized region. Comparing each sub-panel from top to bottom mainly shows the effect from the spread in bubble sizes at a given stage of reionization. Figure~\ref{append_fig:los_examples} illustrates a diversity in bubble sizes and exterior neutral fractions, along with the redshift evolution in the distribution of neutral and ionized gas. The scatter in the exterior neutral fractions can be gauged by comparing the sightline-specific exterior neutral fractions (olive dashed lines in the figure) with the global average values (red dashed lines).

Figure~\ref{append_fig:dw_examples} shows the corresponding DW transmission for each sightline, illustrating how the bubble sizes and exterior neutral fractions impact the DW profiles. At $z=6.1$, for example, the purple and green sightlines show negligible DW absorption, while the blue and orange sightlines show some absorption. This is mostly the result of the much larger bubble sizes in the in green/purple cases ($R=64$ Mpc versus $R=2.0$ Mpc). However, the orange case has slightly more transmission than the blue sightline: even though the orange sightline crosses a larger number of exterior neutral regions, there is a relatively large and nearby neutral region in the case of the blue sightline that leads to slightly more absorption. The semi-analytic calculation explicitly accounts for the size of the first ionized regions (modulo sensitivity to the threshold choice), but does not account for variations in the external neutral fractions.

The $z=8.3$ examples illustrate similar trends, with the orange and blue sightlines showing more absorption than the green and purple cases, owing mostly to the different bubble sizes in these scenarios ($R = 0.37$ Mpc versus $R=11$ Mpc). Here, in three cases the exact DW profile shows more absorption than the semi-analytic calculation (blue, green, and purple), while in the orange scenario the exact DW gives more transmission. The differences at fixed bubble size reflect the fluctuations in the exterior neutral fraction across the different lines of sight. 
Finally, the $z=12.1$ examples show smaller variations that are almost entirely the result of fluctuations in the size of the first ionized region. 

In this context, it is also important to note that the transmission coefficients are only sensitive to the DW profiles quite close to line center. The precise wavelength offsets that are important for the transmission coefficient calculations depend on the linewidths and outflows (e.g. see Figure~\ref{fig:lya_profiles}), but only $\delta_\lambda \lesssim 2-3 \times  v_{\rm circ}/c$ or so are relevant. Especially in this regime, the semi-analytic approximation provides a good description; the exterior neutral fraction variations are of sub-dominant importance.

\section{EW Distribution Derivations}
\label{append_sec:EW_dist}

Here, we give derivations of our conditional EW distribution formulas (Eqns. \ref{eq:EW_dist_cond} and \ref{eq:EW_in_infer}). The key to these derivations is simply to note that $\mathrm{EW}_{\mathrm{obs}} = \mathcal{T}_\alpha \times \mathrm{EW}_{\rm in}$. Therefore, the conditional distribution for the intrinsic EW distribution, given the observed EW follows as:
\begin{equation}
\mathrm{P}(\mathrm{EW}_{\mathrm{in}}| \mathrm{EW}_{\mathrm{obs}}) = \delta_D \left[\mathrm{EW}_{\mathrm{in}} - \frac{\mathrm{EW}_{\mathrm{obs}}}{\mathcal{T}_{\alpha, z}} \right],
\end{equation}
where $\delta_D$ is a Dirac delta function, and $\mathcal{T}_{\alpha, z}$ is the transmission at redshift $z$. Here, we leave the conditioning on $M_{\rm UV}$ implicit for brevity of notation.
By the same deterministic relation, the conditional distribution for $\mathrm{EW}_{\mathrm{obs}}$ given $\mathrm{EW}_{\mathrm{in}}$ is:
\begin{equation}
\mathrm{P}(\mathrm{EW}_{\mathrm{obs}}| \mathrm{EW}_{\mathrm{in}}) = \delta_D \left[\mathrm{EW}_{\mathrm{obs}} - \mathrm{EW}_{\mathrm{in}} \times \mathcal{T}_{\alpha, z} \right].
\end{equation}

For starters, we can use the first relation above to derive Eqn.~\ref{eq:EW_in_infer}, our inference of the intrinsic conditional EW distribution given the observed one at $z=6$ and the IGM transmission statistics at this redshift from \textsc{Thesan}. Specifically, this gives:
\begin{align}
    \mathrm{P}_{\rm in}(\mathrm{EW}_{\rm{in}}|M_{\mathrm{UV}}) &= \int \int d\mathcal{T}_{\alpha, 6} \, d\mathrm{EW_6} \, \mathrm{P}_{\alpha}(\mathcal{T}_{\alpha, 6} | M_{\mathrm{UV, 6}}) \nonumber \\
    &\hspace{1.25 cm} \times \mathrm{P}_{6}(\mathrm{EW_6} | M_{\mathrm{UV, 6}}) \nonumber \\
    &\hspace{1.25cm} \times \delta_{D} \left[\mathrm{EW}_{\rm{in}} - \frac{\mathrm{EW}_{6}}{\mathcal{T}_{\alpha, 6}} \right] \nonumber \\
    &= \int_0^1 d\mathcal{T}_{\alpha, 6} \, \mathcal{T}_{\alpha, 6} \, \mathrm{P}_{\alpha}(\mathcal{T}_{\alpha, 6} | M_{\mathrm{UV, 6}}) \nonumber \\
    &\hspace{1.25cm} \times \mathrm{P}_{6}(\mathrm{EW_{\rm{in}}\mathcal{T}_{\alpha, 6}} | M_{\mathrm{UV, 6}}),
    \label{append_eqn:EW_in}
\end{align}
where $\mathrm{P}_{\alpha}(\mathcal{T}_{\alpha, 6} | M_{\mathrm{UV, 6}})$ denotes the Ly-$\alpha$ transmission coefficient distribution at $z = 6$ ($\mathcal{T}_{\alpha, 6}$). Note that this is implicitly marginalized over bubble sizes and the other physical properties that shape the transmission coefficient distributions.
The quantity $\mathrm{P}_{6}(\mathrm{EW_{\rm{in}}\mathcal{T}_{\alpha, 6}} | M_{\mathrm{UV, 6}})$ is the observed conditional EW distribution, with $\mathrm{EW_{6}} =  \mathrm{EW_{\rm{in}} \times \mathcal{T}_{\alpha, 6}}$. Our fiducial analysis uses the empirical $z=6$ lognormal conditional EW distribution results from  \citet{Tang_2024b}, as summarized in Eqn.~\ref{eq:psix_logn} and Table~\ref{tab:tang24}. The final line of Eqn.~\ref{append_eqn:EW_in} gives the desired result, equivalent to Eqn.~\ref{eq:EW_in_infer} in the body of the text. 

After determining the resulting intrinsic EW distribution, we can derive the observed conditional EW distribution at $z > 6$, assuming that the intrinsic distribution is redshift invariant. Here, we marginalize over the intrinsic EW distribution (while the previous expression marginalized over the observed distribution at $z=6$). This yields:
\begin{align}
    \mathrm{P}_{z}(\mathrm{EW}_{\mathrm{obs}} | M_{\mathrm{UV, z}}) &= \int \int d\mathcal{T}_{\alpha, z} \, d\mathrm{EW_{\rm{in}}} \, \mathrm{P}_{\alpha}(\mathcal{T}_{\alpha, z}| M_{\mathrm{UV},z}) \nonumber \\
    &\hspace{1.25cm} \times \mathrm{P}_{\rm in}(\mathrm{EW_{\rm{in}}} | M_{\mathrm{UV}}) \nonumber \\
    &\hspace{1.25cm} \times \mathrm{P}(\mathrm{EW}_{\mathrm{obs}} | \mathrm{EW}_\mathrm{in}) \nonumber \\
    &= \int \int d\mathcal{T}_{\alpha, z} \, d\mathrm{EW_{\rm{in}}} \, \mathrm{P}_{\alpha}(\mathcal{T}_{\alpha, z}| M_{\mathrm{UV},z}) \nonumber \\
    &\hspace{1.25cm} \times \mathrm{P}_{\rm in}(\mathrm{EW_{\rm{in}}} | M_{\mathrm{UV}}) \nonumber \\
    &\hspace{1.25cm} \times \delta_{D} \left[\mathrm{EW}_{\mathrm{obs}} - \mathrm{EW}_{\rm{in}}\mathcal{T}_{\alpha, z} \right] \nonumber \\
    &= \int_0^1 d\mathcal{T}_{\alpha, z} \, \frac{1}{\mathcal{T}_{\alpha, z}} \, \mathrm{P}_{\alpha}(\mathcal{T}_{\alpha, z}|M_{\mathrm{UV}, z}) \nonumber \\
    &\hspace{1.25cm} \times \mathrm{P}_{\rm in}\left(\mathrm{EW}_{\mathrm{in}} = \frac{\mathrm{EW}_{\mathrm{obs}}}{\mathcal{T}_{\alpha, z}} \, \middle\vert  \, M_{\mathrm{UV}} \right),
    \label{append_eqn:EW_obs}
\end{align}
where $\mathrm{P}_{\alpha}(\mathcal{T}_{\alpha, z}|M_{\mathrm{UV}, z})$ is the distribution of the Ly-$\alpha$ transmission coefficients at redshift $z$ ($\mathcal{T}_{\alpha, z}$). The final expression above is equivalent to Eqn.~\ref{eq:EW_dist_cond}.


\bibliography{references}{}
\bibliographystyle{aasjournalv7}



\end{document}